\documentclass[11pt,a4paper,final]{article}

\usepackage{amsfonts,amsmath,amssymb,theorem,graphicx,hyperref,color,natbib,bbm,url}

\usepackage{mathrsfs} 
\usepackage{multirow}
\usepackage{color}
\usepackage{xcolor}
\usepackage{caption}
\usepackage{subcaption}
\usepackage{latexsym}
\usepackage{nicefrac}
\usepackage{verbatim}
\usepackage{epstopdf} 
\usepackage{multirow}
\usepackage{algorithmic, algorithm}
\usepackage{epsfig,pstricks}
\newtheorem{condition}{Condition}

\newtheorem{theorem}{Theorem}
\newtheorem{lemma}{Lemma}
\newtheorem{corollary}{Corollary}

\title{Tree based credible set estimation}
\author{
  Jeong Eun.~Lee  \\
  Department of Statistics\\
  University of Auckland\\
  \texttt{kate.lee@auckland.ac.nz}  
 \vspace{.3cm}
 \\
 Geoff K.~Nicholls \\
  Department of Statistics\\
  University of Oxford\\
  \texttt{nicholls@stats.ox.ac.uk} \\
}
\def\Xb{\pmb{X}}
\def\xb{\pmb{x}}
\def\Hr{{\rm H}}
\def\hr{h}
\def\fu{q}
\def\Zb{{\pmb Z}}
\def\zb{{\pmb z}}
\def\yobs{{\pmb y_{obs}}}
\def\tip{{\ \stackrel{P}{\rightarrow}\ }}
\def\yb{{\pmb y}}
\def\Yb{{\pmb Y}}
\def\bth{{\pmb \theta}}
\def\bTh{{\pmb \Theta}}
\def\sb{{\pmb s}}
\def\wsb{{\tilde{\pmb s}}}
\def\ub{{\pmb u}}
\def\vb{{\pmb v}}
\def\ab{{\pmb a}}
\def\bb{{\pmb b}}
\usepackage{mathtools}

\begin{document}
	\maketitle
		
		\subsection*{Abstract}	
			
			Estimating a joint Highest Posterior Density credible set for a multivariate posterior density is challenging as dimension gets larger. Credible intervals for univariate marginals are usually presented for ease of computation and visualisation. There are often two layers of approximation, as we may need to compute a credible set for a target density which is itself only an approximation to the true posterior density. We obtain joint Highest Posterior Density credible sets for density estimation trees given by \citet{Li2016} approximating a density truncated to a compact subset of $\mathbb{R}^d$ as this is preferred to a copula construction. These trees approximate a joint posterior distribution from posterior samples using a piecewise constant function defined by sequential binary splits. 
			We use a consistent estimator to measure of the symmetric difference between our credible set estimate and the true HPD set of the target density  samples. This quality measure can be computed without the need to know the true set. We show how the true-posterior-coverage of an approximate credible set estimated for an approximate target density may be estimated in doubly intractable cases where posterior samples are not available. We illustrate our methods with simulation studies and find that our estimator is competitive with existing methods.
	
	\section{Introduction}
	
	Bayesian credible sets are convenient summaries of parameter location and uncertainty. However, Highest Posterior Density (HPD) credible sets for univariate marginal posterior distributions are typically reported, estimated using the methods of \citet{chen99} from tractable densities and Monte-Carlo samples. Joint multi-dimensional credible sets are less frequently used \citep{besag1995,held2004,Krivobokova2010,sorbye2011}, and although some compromise must be made in representing them on the page, this may be worth the effort, as collections of univariate marginals can mislead, and they can be used to extract a wealth of information about the structure of high dimensional posterior distributions.
	
	Consider a joint credible set with probability mass $\alpha$ (for example, $\alpha=0.9$). Many HPD-set-estimation procedures applied to date are expressed as a product of some class of univariate credible interval \citep{besag1995} or set \citep{sorbye2011}. Univariate marginal sets of coverage $\alpha_{uni}$ for each component are nested and grow with increasing $\alpha_{univ}$. If the target distribution is continuous, then there exists a $\alpha_{uni}$-value, $\alpha_{uni}^*\ge\alpha$ say, such that the multivariate random variable falls within the product space of the marginal credible sets with probability $\alpha$. Approaches differ in how $\alpha_{uni}^*$ is estimated. \citet{besag1995} take empirical marginal quantiles to form joint equal-tail intervals, in their Bayesian work \citet{Krivobokova2010} scale empirical univariate quantiles and \citet{held2004} uses empirical quantiles of robust Rao-Blackwellised posterior density estimates. In contrast \citet{sorbye2011} forms a tractable mixture-of-normals approximation, estimates HPD sets for marginals and forms the joint credible set as the product of these marginal sets. 
	
	More general multivariate HPD credible sets can be estimated from the level sets of a joint density estimate. A common level set estimator is the plug-in estimator using a density estimate. Asymptotic properties such as consistency and rates of convergence have been studied \citep{Tsybakov1997,Cadre2006}. \citet{Mammen2013} gives examples of statistical inference for level sets. Existing density estimation algorithms can be categorized into parametric, semi-parametric and nonparametric approaches. One widely-adopted parametric or semi-parametric density estimation algorithm fits a Gaussian Mixture Model (GMM) \citep{Pearson1894, Aitkin80, McLachlan08, Wang15} to the target. Existing nonparametric methods for density estimation include histograms \citep{Scott79}, frequency polygons \citep{Scott85}, Kernel Density Estimation \citep{Scott77, SILVERMAN86, chen17}, Splines \citep{STONE94}, the Mondrian process \citep{roy08} and neural network-based estimators \citep{NIPS1998_1624,LAROCHELLE11,NIPS2017_6828}. Level-set approaches make it easy to evaluate the point-in-HPD-set indicator function (it equals the indicator for the density estimate at the point to exceed the density estimate at the level set). Topological information about the set is also available. 
	
	Methods based on level sets of KDE's for densities are particularly well developed and these apply straightforwardly to HPD set estimation. Recent work gives estimation procedures, convergence of set estimates and set error estimates for measures of the symmetric difference between the estimate and target set \citep{baillo00,mason09} with convergence in Hausdorff loss and bootstrap confidence regions for sets \citep{chen17}. We use similar set error estimates of symmetric difference although, as \citet{baillo06} note in connection with Hausdorff convergence of support-estimation with their histogram method, a loss sensitive to set shape will sometimes be preferable. Level set trees (\citet{Klemela2004} and \citet{stuetzle10} and earlier work by the same authors) give a useful summary of the mode structure, mass and magnitude as the level set threshold is varied, and further tools for visualisation of relations between level sets using graphs on MDS projections are given in \citet{chen17}.

	If all we need is point-in-HPD-set evaluation, then it is not always necessary to make a density estimate. \citet{held2004} exploits this, using an observation due to \citet{box65}. If we have $n$ samples from the target density, then the $\lfloor n(1-\alpha)\rfloor$ order statistic of the unnormalised target density at the samples is a consistent estimate for the level set threshold of the unnormalised density. We make use of this when the posterior can be evaluated up to a constant. However, density ratio evaluation is not possible when the posterior is doubly intractable. We show how to estimate and calibrate HPD sets in this setting. Point-in-HPD-set queries have been suggested as a way of making a Bayesian Hypothesis test for the true parameter to be located at the test point, by analogy with the Frequentist relation between $p$-values and confidence intervals. However, although a decision-theoretic basis in terms of a loss does exist, the loss involved will not in general match any real user loss \citep{thulin14}, so that point-in-HPD-set evaluation is mainly useful as an exploratory tool to locate posterior probability mass. 
	
	We found the HPD-set estimates formed from level-sets of KDE's to be efficient statistically, but demanding to compute. We seek a similarly principled method for joint HPD-set estimation that scales straightforwardly to large sample sizes ($10^6$) and moderate dimensions (say, twenty). We form a non-parametric density estimate from MCMC-samples as a first stage using Density Estimation Trees (DET, \citet{RamGray2011}), and in particular the $D^*$-Partition (DSP) DET-estimator given in \citet{Li2016}, with axis-parallel splits and a piece-wise constant density estimate. \citet{Li2016} show consistency for probability mass estimates on hyper-rectangular sets using error bounds based on the star-discrepancy \citep{Niederreiter92} of the partition defined by the tree. Their experiments show that the DSP estimate itself (ie without forming the HPD set) is fast and accurate, competitive with KDE for statistical efficiency and far faster.
	We reproduce their outline of relevant theory in Section~\ref{sec:dentree}. An estimate of the joint HPD credible set is defined in Section~\ref{sec:hpdest} as a union of the highest density leaf-sets. This joint credible set can differ a great deal from the product of marginal HPD sets. 
	
	
	 Our implementation of the DSP DET-estimator in \citet{Li2016} did not exploit the computational tricks suggested in that paper, but nevertheless generated well-calibrated HPD-set estimates with acceptable runtimes.  \citet{RamGray2011} outline DET-construction based on the classical tree learning algorithm of \citet{Breiman84} with moditifications for the density estimation setting \citep{Breiman84}. \citet{LuJiangWong2013} implicitly covers binary partitioning in a Bayesian setting with convergence rate to the posterior given in \citet{LiuLiWong2017}. Lately \citet{Wu2018} adapted random forest methods to the Bayesian sequential partitions \citet{LuJiangWong2013}. Simulation studies of these piecewise-constant density estimators show that this approach is often practical for moderate dimensional densities. 
	 
	 When we construct a candidate HPD-set it is important to have some measure of its quality. When target density ratios can be evaluated, we measure the loss, which we define in Section~\ref{sec:loss} as the posterior probability mass on the symmetric difference between the set-estimate and the true HPD set. In our setting this estimator follows from \cite{box65}: as the level-set threshold estimated on test samples converges it identifies test samples within and outside the true HPD set and these can be used to estimate the probability mass in the symmetric difference. We present the estimator in Theorem~\ref{thm5} in order to be clear how it applies in our setting. However, this kind of set-error estimation is not new in the set-estimation literature. Similar methods can be found in \cite{baillo00} with extensions leading up to \cite{mason09} and related ideas in \cite{chen17}. When the loss can be estimated we use it to select a ``bandwidth'' hyperparameter of the DET estimate. When target density ratios cannot be evaluated (ie Bayesian inference with a doubly intractable posterior), we give a loss estimate based on a lower bound. We show in Section~\ref{sec:converge} that, if the HPD-set estimate converges, then the loss converges to this bound. However, convergence is an assumption here, in contrast to the KDE literature, where convergence of level-set estimates is known. 
	 
	 However, in the Bayesian doubly-intractable setting, the samples on which set estimates are based often come from some distribution which is only an approximation to the real target posterior, for example, in Variational Bayes and Approximate Bayesian Computation. In this setting set convergence is secondary as it would in any case converge to the ``wrong'' set. Some ``external'' measure of set-estimate error is needed. Recently \citet{Lee2019}, \citet{Xing2019} and \citet{Xing20} gave calibration procedures for approximate Bayesian inference which allow us to calibrate the HPD set estimates we form on the true posterior. Their studies are limited to two dimensional marginal posterior distributions. We explain how to calibrate multivariate credible sets for exact and approximate multivariate posterior densities Section~\ref{sec:calibtheory}. In Section~\ref{sec:sim}, our credible set estimators are examined numerically and calibrated on simulated and real datasets. The paper ends with a conclusion in Section~\ref{sec:conclusion}. An R implementation of the density estimation tree and proposed credible set estimators with examples can be found in \url{https://github.com/klee61/TCS}.
	

	

 	\section{Density Estimation}\label{sec:dentree}
 	
 	The material in this section follows \citet{Li2016} closely and is included to make the presentation self-contained. The main variation is that we avoid methods based on estimating a copula in $[0,1]^d$, used by those authors in their practical density-estimation examples, and replace it with direct estimation of the density, restricted, if necessary, to a ``truncation'' set chosen automatically. We explain why we do things this way in Section~\ref{sec:notcopula}.
 	
 	In the following the distribution $F$ will be taken to be a posterior distribution arising from some Bayesian analysis. 
 	We will be interested in computing and representing an HPD credible set for $F$. 
 	
 	Let $\Xb=(X_{1},X_{2},\ldots,X_{d})$ be a continuous multivariate random variable with $d$-dimensional sample space $\Omega\subseteq\mathbb{R}^d$, sigma-algebra $\mathcal{B}$ and probability distribution $F:\mathcal{B}\rightarrow [0,1]$.
 	Let $\xb=(x_{1},\ldots,x_{d})$, $\xb\in\Omega$ be a realisation of $\Xb$. 
 	We assume $F$ has a density $f$, $dF(\xb)=f(\xb)d\xb$ with respect to Lebesgue measure $d\xb$ in $\Omega$. 
 	Suppose that for $j=1,\ldots,n$ we have $\Xb^{(j)}\sim F$ independent and identically distributed (iid) with $\Xb^{(j)}=\xb^{(j)}$ a realisation.
 	Let $\Xb^{(1:n)}=\{\Xb^{(j)}\}_{j=1,\ldots, n}$ and ${\xb}^{(1:n)}=\{\xb^{(j)}\}_{j=1,\ldots, n}$.
 	
 	\subsection{Density estimation in a fixed truncation set}\label{sec:truncate} 
 	 When the set $\Omega$ is not bounded it must for our purpose be truncated. HPD-set estimation can be done without loss if the coverage of the truncation set is known and the HPD set is a subset. For truncation-level $\alpha\le p\le 1$ let $\Omega^{(p)}$ denote some fixed truncated set with $F(\Omega^{(p)})=p$. 
 	 We take $\Omega^{(p)}$ to be hyper-rectangular, 
 	 \begin{equation}\label{eq:omp}
 	 \Omega^{(p)}=[ \omega^{p,-},\omega^{p,+}],    
 	 \end{equation} 
 	 with  $\omega^{p,\pm}=(\omega^{p,\pm}_{1},\ldots,\omega^{p,\pm}_{d})$ and finite boundaries $-\infty<\omega^{p,-}_i<\omega^{p,+}_i<\infty,\ i=1,\ldots,d$. 
 	 Denote by $\mathcal{B}^{(p)}$ the Borel $\sigma$-algebra of sets in $\Omega^{(p)}$ and let $F^{(p)}$ give the conditional distribution of $\Xb|\Xb\in\Omega^{(p)}$ with density $f^{(p)}$.
 	 
 	 Taking $F$ to be a posterior density with parameter vector $\xb\in\Omega$, we define the Highest Posterior Density set (HPD set) with coverage $0\le\alpha\le 1$ to be the set 
 	 \begin{equation}\label{eq:hpd_def1}
 	   Q^\alpha=\{\xb \in \Omega: f(\xb)>\gamma\}  
 	 \end{equation} 
 	 with $\gamma$ chosen so that $F(Q^\alpha)=\alpha$ (in our notation $\alpha$ is typically a number close to one). We assume this set is unique. We refer below to $Q^\alpha$ as a ``level-$\alpha$'' HPD set or a set ``with coverage $\alpha$''.
 	 \begin{condition} \label{cond3}
 	 Let $Q^\alpha$ denote the true level-$\alpha$ HPD set for $F$. The truncation set $\Omega^{(p)}$ defined in Eqn.~\ref{eq:omp} satisfies $F(\Omega^{(p)})=p$ and $Q^\alpha\subset \Omega^{(p)}$, so that the true HPD set is contained in $\Omega^{(p)}$.
 	 \end{condition}
 	 Under Condition~\ref{cond3} the level-$\alpha$ HPD set of $F$ is the level-$\alpha/p$ HPD set of $F^{(p)}$.
 	Care is needed to ensure $Q^\alpha$ is contained in $\Omega^{(p)}$.
 	 We will later replace $\Omega^{(p)}$ with an estimated set $\hat\Omega^{(p)}$, a step we justify in Section~\ref{sec:estimatetruncate}. In the algorithm we implemented we took $\hat\Omega^{(p)}$ to be the smallest hyper-rectangular set containing all points $\xb^{(1:n)}$ in the realised sample. This is an estimate of a set $\Omega^{(p)}$ with $p=1-2/n'$ for $n'$ equal the number of samples in the training set we used. When we discuss asymptotics it is understood that this $p$ is then fixed as $n$ increases. A user might alternatively take a fixed set $\Omega^{(p)}$ with unknown coverage $p$ which they are confident contains $Q^\alpha$. In either case, if $p$ is this close to one then the associated error in the estimated HPD set is negligible compared to the Monte Carlo error in the set estimate and so although we account for this in our analysis, we ignored it in our implementation. 
 	 
	We adopt the methods of \citet{Li2016}. The goal of density estimation is to find a binary partition 
	 $\Lambda=\{\Lambda_1,\ldots,\Lambda_K\}$ of $\Omega^{(p)}$ into hyper-rectanglar cells such that the density $f^{(p)}$ is well approximated by a constant in each cell. 
	For $k=1,\ldots,K$ let $\Lambda_k=\{\xb\in \Omega^{(p)}: u_{k,i}\le x_i\le v_{k,i}\}$ be a hyper-rectangular cell with volume $|\Lambda_k|$ in a $K$-set partition (up to sets of $F$-measure zero) of $\Omega^{(p)}$. 
	The partition sets $\Lambda_k$ correspond to the leaves of a tree computed by the tree-building algorithm, the DSP Algorithm of \citet{Li2016}, reproduced as Algorithm~\ref{alg:detcore} in Appendix. 
	Algorithm~\ref{alg:detcore} takes as input a sample realisation $\Xb^{(1:n)}=\xb^{(1:n)}$ and target space $\Omega^{(p)}$ and outputs the partition $\Lambda=\Lambda(\xb^{(1:n)};\Omega^{(p)})$. 
	
	The density estimate is determined from the sample points $\xb^{(1:n)}$ and partition $\Lambda$ in the following way.
	For $k=1,\ldots,K$, let 
	\[
	n_k=\sum_{j=1}^n \mathbbm{1}(\xb^{(j)}\in \Lambda_k)
	\] count the samples in set $\Lambda_k$. 
	The tree-based density estimator in $\Omega^{(p)}$ is $\tilde f^{(p)}_n$ where, for $\xb\in\Omega^{(p)}$, 
	\begin{align}\label{eq:ftilde}
	    \tilde f^{(p)}_n(\xb; \xb^{(1:n)},\Omega^{(p)})&=\frac{1}{N}\sum^K_{k=1} \frac{n_k}{|\Lambda_k|}  \mathbbm{1}(\xb\in \Lambda_k) \, ,\\
	    \intertext{and $N=N(\xb^{(1:n)};\Omega^{(p)})$ counts the samples in $\Omega^{(p)}$,} \label{eq:N}
	    N(\xb^{(1:n)};\Omega^{(p)})&=\sum_{j=1}^n \mathbbm{1}(\xb^{(j)}\in\Omega^{(p)}).
	\end{align}
	  
	 \citet{Li2016} construct the tree using the ``star-discrepancy'' to measure the degree of uniformity of points in a sub-rectangle. This measure is used in forming bounds for quasi-Monte Carlo integration error (for example, \citet{Niederreiter92}).
	 Given $n^*$ points $\wsb^{(1)},...,\wsb^{(n^*)}$ in $[0,1]^d$, the star discrepancy is
	 \begin{equation}\label{eq:dstar}
	     D^* ( \{\wsb^{(j)}\}^{n^*}_{j=1})  = {\rm sup}_{\pmb{a}\in[0,1]^d} \left |
	 \dfrac{1}{n^*} \sum^{n^*}_{j=1} \mathbbm{1} (\wsb^{(j)}\in [ \pmb{0},\pmb{a} ) ) -\prod^d_{i=1} a_i \right|. 
	 \end{equation}
	 The supremum is taken over all $d$-dimensional sub-rectangles $[\pmb{0},\pmb a)$ with the origin at a corner. In our implementation of the DSP Algorithm, Algorithm~\ref{alg:detcore}, the supremum is approximated by maximising over a finite lattice of values. 
	 
	 Let $s(\cdot;\Lambda_k): \Lambda_k\rightarrow [0,1]^d$ be an affine transformation mapping points $\xb$ in hypercube $\Lambda_k$ by
	 \begin{equation}\label{eq:affine_map_s}
	     s(\xb;\Lambda_k)=\left(\frac{ x_1 - u_{k,1}}{v_{k,1}-u_{k,1}},...,\frac{ x_d - u_{k,d}}{v_{k,d}-u_{k,d}}\right),
	     \end{equation}
	 so that $s(\xb;\Lambda_k)\in [0,1]^d$. For $k=1,\ldots,K$ and $j=1,\ldots,n_k$ denote by $\xb^{(k,j)}\in \Lambda_k$ the locations of the $n_k$ samples in $\xb^{(1:n)}$ and that are in $\Lambda_k$, so that $D^*(\{s({\xb}^{(k,j)};\Lambda_k)\}_{j=1}^{n_k})$ is a standardised measure of the uniformity of points in $\Lambda_k$.
	 
	 Denote by $V(g)$ the {\it variation}, in the sense of Hardy and Krause, of any given function $g:\Omega^{(p)}\rightarrow \mathbb{R}$. We refer the reader to \citet{Niederreiter78} and \citet{Owen2005} for the definition of $V(g)$ and further discussion. The following Theorem, which follows from the Koksma-Hlwaka inequality, is Theorem~2 in \cite{Li2016}, where it is set in $[0,1]^d$. We work in the compact hyper-rectangular set $\Omega^{(p)}$. However the transformation between $\Omega^{(p)}$ and $[0,1]^d$ is a linear, strictly monotone and invertible function and so the variation is unchanged \citep{Owen2005}. 
	 	\begin{theorem}\label{thm3} \citep{Li2016}
		Let $g$ be a function defined on $\Omega^{(p)}$ with bounded variation $V(g)$. Let $\Lambda=\{\Lambda_1,...,\Lambda_K\}$ be a binary partition of $\Omega^{(p)}$. Let a realisation $\Xb^{(j)}=\xb^{(j)}\in \Omega,\ j=1,\ldots n$ (sorted so that $\xb^{(1:N)}\in\Omega^{(p)}$ are the $N$ points in $\Omega^{(p)}$), and $\tau>0$ be given. If for each $k=1,\ldots,K$ 
		\begin{equation}\label{eq:bound_on_Dstar}
		    D^*(\{s({\xb}^{(k,j)};\Lambda_k)\}_{j=1}^{n_k}) \leq \frac{\tau\sqrt{N}}{n_k},
		\end{equation}   
	then the absolute difference between the expectation of $g$ computed in the density estimate $\tilde f^{(p)}_n(\xb;\xb^{(1:n)},\Omega^{(p)})$ (in Eqn.~\ref{eq:ftilde}) and the average, $\bar g(\xb^{(1:N)})={N}^{-1}\sum_j g(\xb^{(j)})$ is uniformly bounded,
	 \begin{equation}\label{eq:outcome_bound_thm3}
	     \left| \displaystyle\int_{\Omega^{(p)}} g(\xb)\tilde{f}^{(p)}_n(\xb;\xb^{(1:N)},\Omega^{(p)}) d\xb - \bar g(\xb^{(1:N)})\right|  \leq \dfrac{\tau}{\sqrt{N}}V(g).
	 \end{equation}
	\end{theorem}
	The parameter $\tau$ in Theorem~\ref{thm3} plays the role of a bandwidth.
	If $\tau$ is relatively larger then the $D^*$-condition in Eqn.~\ref{eq:bound_on_Dstar} may be satisfied by a relatively coarser partition, as relatively larger deviations from uniformity are allowed. 
	
	The condition given in Eqn.~\ref{eq:bound_on_Dstar} is exactly the bound taken by \citet{Li2016} in their DSP algorithm. Those authors actually demonstrate Eqn.~\ref{eq:outcome_bound_thm3} holds under a different bound which we do not detail. However, it is straightforward to check that the bound verified by \citet{Li2016} in their DSP Algorithm, and taken in Eqn.~\ref{eq:bound_on_Dstar}, can be substituted in the proof given by \citet{Li2016} and leads to Eqn.~\ref{eq:outcome_bound_thm3}.
	
	
	Algorithm~\ref{alg:detcore} in the Appendix includes an extra step which we now explain. 
	At Step 2 we map the samples $\xb^{(1:N)}\in \Omega^{(p)}$ to points $\sb^{(1:N)}\in [0,1]^d$, using the affine transformation $\sb^{(1:N)}=s(\xb^{(1:N)};\Omega^{(p)})$ defined in Eqn.~\ref{eq:affine_map_s}, and then carry out density estimation in $[0,1]^d$ using the samples $\sb^{(1:N)}$. This is what we actually do on the computer as we are working in a hyper-rectangular set $\Omega^{(p)}$ and it is convenient to standardise. Algorithm~\ref{alg:detcore} finds a good partition $\Delta=\{\Delta_k\}_{k=1}^K$ in $[0,1]^d$ which we map back as sets $\Lambda_k=s^{-1}(\Delta_k; \Omega^{(p)}),\ k=1,...,K$ at Step 22 to get the corresponding partition $\Lambda=\{\Lambda_k\}_{k=1}^K$ of $\Omega^{(p)}$. The density estimate $\tilde f_n^{(p)}$ given at the end of Algorithm~\ref{alg:detcore}, which is identical to Eqn.~\ref{eq:ftilde}, takes account of the Jacobian of the transformations $s$ into $[0,1]^d$ and $s^{-1}$ back to $\Omega^{(p)}$. These are respectively $|\Omega^{(p)}|^{-1}$ and $|\Omega^{(p)}|$ and cancel. 
	
	If the pair of transformations in and out of $[0,1]^d$ in lines 2 and 22 are omitted we get the same final partition $\Lambda$ of $\Omega^{(p)}$ either way, because the composition of affine mappings
	 $\sb=s(\xb;\Omega^{(p)})$, $\Delta_k=s(\Lambda_k;\Omega^{(p)})$ and $\wsb=s(\sb;\Delta_k)$ is equal to the single affine mapping $\wsb=s(\xb;\Lambda_k)$. In Theorem~\ref{thm3}, $D^*(\{s({\xb}^{(k,j)};\Lambda_k)\}_{j=1}^{n_k})$ is calculated from the partition $\Lambda$ of $\Omega^{(p)}$. In Algorithm~\ref{alg:detcore}, $D^*(\{s({\sb}^{(k,j)};\Delta_k)\}_{j=1}^{n_k})$ is calculated from the partition $\Delta_k$ of $[0,1]^d$. However, these $D^*$ values are equal because the $D^*$ values are computed on the same vectors $\{\wsb^{(k,j)}\}_{j=1}^{n_k},\ k=1,...,K$, whether mapped directly from $\Omega^{(p)}$ or via $[0,1]^d$.
	
	We consider now convergence of probability mass on sets in $\Omega^{(p)}$. For $i=1,\ldots,d$ let $\omega^{p,-}_i\le u_{i}\le v_{i}\le\omega^{p,+}_i$ give the bounds along dimension $i$ of $\Omega^{(p)}$ in Eqn.~\ref{eq:omp}, of a hyper-rectangular cell in $\Omega^{(p)}$,
	\begin{equation}\label{eq:hrs}
	    \Hr=\{\xb\in\Omega^{(p)}: u_{i}\le x_i\le v_{i},\ i=1,\ldots,d\}.
	\end{equation}
	Let $\mathcal{H}^{(p)}$ be the set of all hyper-rectangular subsets of $\Omega^{(p)}$ of the form given in Eqn.~\ref{eq:hrs}. Let $\hr(\xb;\Hr)=\mathbb{I}_{\xb\in \Hr}$ and $\bar\hr(\xb^{(1:n)})=\frac{1}{n}\sum_{j=1}^n \hr(\xb^{(j)}; \Hr)$. For $\Hr\in \mathcal{H}^{(p)}$ let
	\[
	\tilde F^{(p)}_n(\Hr)=\int_{\Omega^{(p)}} \hr(\xb;\Hr) \tilde f^{(p)}_n(\xb; \Xb^{(1:n)})\, d\xb
	\]
	give the approximate distribution on hyper-rectangular sets.
	The following corollary is a simple restatement of Corollary~5 of \citet{Li2016} for the case where the number of points $N$ is random but $np/N$ converges almost surely to one. For example, $\Xb^{(j)},\ j=1,\ldots,n$ may be simulated using a Markov chain targeting $F$.
	
	\begin{corollary}\label{cor4} Let a hyper-rectangular set $\Hr\in \mathcal{H}^{(p)}$ be given. For each $n>0$, let $\Xb^{(1:n)}=(\Xb^{(1)},\ldots,\Xb^{(n)})$ be a set of $n$ $d$-dimensional random variables satisfying $\bar\hr(\Xb^{(1:n)})\stackrel{a.s.}{\rightarrow} F(\Hr)$. If $\tilde f^{(p)}_n(\xb; \Xb^{(1:n)},\Omega^{(p)})$ is the approximation given in Eqn.~\ref{eq:ftilde} in terms of the random partition $\Lambda(\Xb^{(1:n)};\Omega^{(p)})$ output by DSP Algorithm~\ref{alg:detcore}, then
		\[
		\tilde F^{(p)}_n(\Hr)\stackrel{a.s.}{\rightarrow}  F^{(p)}(\Hr)\, ,
		\]
		where $F^{(p)}(\Hr)=F(\Hr)/p$ is the conditional distribution of $\Xb\sim F$ given $\Xb\in \Omega^{(p)}$.
	\end{corollary}
	Proof: following \citet{Li2016} (and dropping implied arguments from the notation),
	\begin{align}\label{eq:tip}
	\left\vert\, p\int_{\Omega^{(p)}}\hr(\xb;\Hr)\tilde f^{(p)}_n(\xb)\,  d\xb-F(\Hr)\right\vert&= \left\vert\, p\int_{\Omega^{(p)}} \hr(\xb;\Hr)\tilde f^{(p)}_n(\xb) d\xb-\frac{np}{N}\bar\hr+\frac{np}{N}\bar\hr -F(\Hr)\right\vert\nonumber\\
	&\le p\frac{\tau V(\hr)}{\sqrt{N}} +\left\vert \frac{np}{N}\bar\hr- F(\Hr)\right\vert,
	\end{align}
	for $N\ge 1$. Using results from \citet{Owen2005}, \citet{Li2016} show that $V(\hr)$ is finite, so the difference of the first two terms in the first line of Eqn.~\ref{eq:tip} is uniformly bounded over $\Xb^{(1:n)}=\xb^{(1:n)}$ (at each $N$) by $\tau V(\hr)/\sqrt{N}$, by Theorem~\ref{thm3}. The terms in the second line converge almost surely to zero. As \citet{Li2016} note, the requirement that $V(\hr)$ be finite restricts the proof to convergence on sets $\Hr\in\mathcal{H}^{(p)}$.
     [End of Proof] 
     
     Let $\mathcal{C}^{(p)}$ be the set of all subsets of $\Omega^{(p)}$ which can be represented by taking countable unions of hyper-rectangular sets $\Hr_k\in \mathcal{H}^{(p)},\ k\ge 1$. 
   By Corollary~\ref{cor4}, the distribution $\hat F^{(p)}_n$ converges weakly to a distribution $\tilde F^{(p)}_\infty$ coinciding with $F$ on sets in $\mathcal{C}^{(p)}$. This limit distribution has a unique extension to a distribution on Borel sets $\mathcal{B}^{(p)}$, as every open set in $\mathbb{R}^d$ is a countable union of hyper-rectangular sets. The extension is equal to $F$ and so the density of $\tilde F^{(p)}_\infty$ coincides with $f$ except on sets of zero measure.
   This is a statement about the density of the limit, rather than the limit of the density and hence is not of direct use.
    
	 \subsection{Density estimation in an estimated truncation set} \label{sec:estimatetruncate}
	 
	 In practice we would like to estimate the truncation set $\Omega^{(p)}$ in Eqn.~\ref{eq:omp} using the same samples $\Xb^{(1:n)}$ we use to construct the density estimate. There is then a selection process thinning the samples down to those inside the estimated truncation set. When we need to truncate in the examples later in the paper, the boundaries of $\Omega^{(p)}$ are quantiles of the marginal distribution of $F$ on each dimension, and these are estimated using order statistics. However we truncate, we need the conditional distribution of the samples we select to have distribution $F$ within the random truncation set. This requirement is expressed in Condition~\ref{cond2} below, which is satisfied by truncation based on order-statistics.
	 
	  Let $\hat\Omega^{(p)}_n=[\hat \omega^{p,-}_n,\hat\omega^{p,+}_n]$ be an estimate of $\Omega^{(p)}$ given by estimates $\hat \omega^{p,\pm}_n=(\hat \omega^{p,\pm}_{n,1},\ldots,\hat \omega^{p,\pm}_{n,d})$ of the boundaries $\omega^{p,\pm}$ of $\Omega^{(p)}$. These estimates are computed using the realisation $\Xb^{(1:n)}=\xb^{(1:n)}$. For $A,B\in \mathcal{B}$ let
	\begin{equation}\label{eq:sd}
	    A\,\Delta\,  \, B = (A\setminus B) \cup (B\setminus A)
	\end{equation} 
	denote the symmetric difference between $A$ and $B$.
	We assume the estimator $\hat\Omega^{(p)}(\Xb^{(1:n)})$ converges in the following sense. 
	\begin{condition}\label{cond1}
	The estimator $\hat\Omega^{(p)}_n=\hat\Omega^{(p)}(\Xb^{(1:n)})$ satisfies \begin{equation*}
	F(\hat\Omega^{(p)}_n \,\Delta\,  \, \Omega^{(p)})\tip  0
	\end{equation*}
	as $n\rightarrow\infty$. 
	\end{condition}
	
	In the next condition we assume further that the rule for constructing $\hat\Omega^{(p)}_n$ does not impact the distribution of the truncated sample, other than by truncation, so that the distribution of the samples in the subset we keep is the same as the distribution of a new sample conditioned on it being in the estimated truncation set.
	\begin{condition}\label{cond2}
	Let $\mathcal{X}^p_n=\{\Xb\in\Xb^{(1:n)}: \Xb\in \hat\Omega^{(p)}_n\}$ denote the truncated sample set of samples in $\hat\Omega^{(p)}_n$. Let $\Xb\sim F$. We require
	\[
	P(\Xb\in\Hr|\Xb\in \mathcal{X}^p_n)=P(\Xb\in\Hr|\Xb\in \hat\Omega^{(p)}_n),
	\]
	for all hyper-rectangular sets $\Hr\in \mathcal{H}^{(p)}$.
	\end{condition}
	Conditions~\ref{cond1} and \ref{cond2} hold in the examples below as the boundaries of $\hat\Omega^{(p)}_n$ are based on order statistics of samples in $\Xb^{(1:n)}$. Condition~\ref{cond1} holds as the $\lfloor n(1-p)/2\rfloor$ order statistic converges in probability to the $(1-p)/2$-quantile. Condition~\ref{cond2} holds as the distribution of the $k$'th order statistic, $X^{[k]}_i$ say, given the $j$'th, $X^{[j]}_i=c$ say, is for $k>j$ the same as the distribution of the $(k-j)$'th order statistic in a sample of size $n-j$ drawn from $F$ and conditioned on $X_i>c$. 
	
	Under the conditions above, the truncation procedure does not introduce an asymptotic bias when we use the samples which remain to estimate set probability mass.
	\begin{lemma}\label{lem2}
	 Let $\Hr\in \mathcal{H}^{(p)}$ be given. Under Conditions~\ref{cond3}, \ref{cond1} and \ref{cond2}, $F(\hat\Omega^{(p)}_n)\tip p$ and
	\begin{equation}\label{eq:cond-omega-hat}
	P(\Xb\in\Hr|\Xb\in \mathcal{X}^p_n)\tip  F^{(p)}(\Hr).
	\end{equation} 
	\end{lemma}
	Proof: to show that first part observe that
	\[
	F(\hat\Omega^{(p)}_n)+ F(\Omega^{(p)}\setminus\hat\Omega^{(p)}_n)=F(\Omega^{(p)})+ F(\hat\Omega^{(p)}_n\setminus\Omega^{(p)})
	\]
	and apply Conditions~\ref{cond3} and \ref{cond1}. To show Eqn.~\ref{eq:cond-omega-hat}, apply Condition~\ref{cond2} to the left side and consider 
	\[
	P(\Xb\in\Hr|\Xb\in \hat\Omega^{(p)}_n)=F(\Hr\cap\hat\Omega^{(p)}_n)/F(\hat\Omega^{(p)}_n).
	\] 
	Decompose $\Hr$ as
	\[\Hr=(\Hr\cap\hat\Omega^{(p)}_n) \cup (\Hr\cap (\Omega^{(p)}\setminus \hat\Omega^{(p)}_n)),\]
    which holds since $\Hr\subset \Omega^{(p)}$. By Condition~\ref{cond1}, $F(\Hr\cap (\Omega^{(p)}\setminus \hat\Omega^{(p)}_n))\tip 0$ and so
    $F(\Hr\cap\hat\Omega^{(p)}_n)\tip F(\Hr)$ and
     $P(Y\in\Hr|Y\in \hat\Omega^{(p)}_n)\tip F(\Hr)/p$ follows from the first part. [End of Proof]

	Let $\hat f^{(p)}_n(\xb; \Xb^{(1:n)},\hat\Omega^{(p)}_n)$ be given by Eqns~\ref{eq:ftilde} and \ref{eq:N} (with $\hat\Omega^{(p)}_n$ replacing $\Omega^{(p)}$) and let
	 \begin{equation}\label{eq:Fhatp}
	  \hat F^{(p)}_n(\Hr)=
	  \int_{\hat\Omega^{(p)}_n} \hr(\xb)\hat f^{(p)}_n(\xb; \Xb^{(1:n)},\hat\Omega^{(p)}_n) d\xb.
	 \end{equation}
	 The following Corollary of Theorem~\ref{thm3} shows that, when we switch to working in an estimated truncation set $\hat\Omega^{(p)}$, the estimated probability mass on hyper-rectangular sets still converges to the right value, as in Corollary~\ref{cor4}. The convergence is now in probability, as the truncation set convergence in Condition~\ref{cond1} is in probability. 
	 \begin{corollary}\label{cor6} 
	 Under the assumptions of Corollary~\ref{cor4} and Lemma~\ref{lem2},
		\[
		\hat F^{(p)}_n(\Hr)\tip   F^{(p)}(
		 \Hr).
		\]
	\end{corollary}
	Proof: following \citet{Li2016} and the proof of Corollary~\ref{cor4}, writing $\hat f^{(p)}_n(\xb)$ for $\hat f^{(p)}_n(\xb; \Xb^{(1:n)},\hat\Omega^{(p)}_n)$, now with samples distributed as $\Xb|\Xb\in \mathcal{X}^p_n$, the conditions of Theorem~\ref{thm3} are satisfied in the random set $\hat\Omega^{(p)}_n$ so,
	\begin{align}\label{eq:tip2}
	\left\vert\, \hat F^{(p)}_n(\Hr)-\frac{F(\Hr)}{p}\right\vert&= \left\vert\, \int_{\hat\Omega^{(p)}_n} \hr(\xb,\Hr)\hat f^{(p)}_n(\xb) d\xb-\frac{n}{N}\bar\hr+\frac{n}{N}\bar\hr -\frac{F(\Hr)}{p}\right\vert\nonumber\\
	&\le \frac{\tau V(\hr)}{\sqrt{N}} +\left\vert \frac{n}{N}\bar\hr- \frac{F(\Hr)}{p}\right\vert,
	\end{align}
    where $N=N(\Xb^{(1:n)},\hat\Omega^{(p)}_n)$ is defined as in Eqn.~\ref{eq:N}.
    As in Corollary~\ref{cor4}, $V(\hr)$ is finite. Now
    \begin{align*}
        \left\vert \frac{n}{N}\bar\hr- \frac{F(\Hr)}{p}\right\vert&\le \left\vert \frac{1}{N} \sum_{\Xb\in \mathcal{X}^p_n} \mathbbm{1}(\Xb\in \Hr)- P(\Xb\in\Hr|\Xb\in \mathcal{X}^p_n)\right\vert+\left\vert P(\Xb\in\Hr|\Xb\in \mathcal{X}^p_n)- \frac{F(\Hr)}{p}\right\vert
    \end{align*}
    and so by Lemma~\ref{lem2} and the WLLN the RHS of Eqn.~\ref{eq:tip2} converges in probability to zero. [End of Proof]
    
    \subsection{Density estimation using a copula construction}\label{sec:notcopula}
    
    In this section we add some remarks on using a copula construction for tree-based density estimation. It may appear natural to map $\Omega$ to $[0,1]^d$ using the empirical marginal CDF's, compute a density estimate for the copula and map back to $\Omega$. This has the advantage of avoiding the need to truncate unbounded sample spaces, as we do above. 
    
    For $i=1,\ldots,d$ and $x_i\in \mathbb{R}$, denote by $F_i(x_i)=\Pr(X_i\le x_i)$ the continuous marginal CDF for $X_i$ with empirical CDF $\widehat{F}_i(x_i)=\frac{1}{n}\sum^n_{j=1} \mathbbm{1} (x_i^{(j)}\leq x_i)$. 
    Transformation of the components of $\Xb$ with the exact marginal CDF's gives the copula representation
 	 \begin{equation}\label{eq:copula}
 	     F(x_1,\ldots,x_d)=C(F_1(x_1),...,F_d(x_d))
 	 \end{equation}
 	 and corresponding copula density $c(s_1,\ldots,s_d)$ for $s_i\in [0,1],\ i=1,\ldots,d$ of \citet{Sklar1959}. For $j=1,\ldots,n$,
 	let $\wsb^{(j)} = (s_1^{(j)},...,s_d^{(j)})$ with $s_i^{(j)} = \widehat{F}_i(x^{(j)}_i)$ for $i=1,\ldots,d$ and let $\wsb^{(1:n)}=(\wsb^{(1)},\ldots,\wsb^{(n)})$. 
 	Denote by $\tilde c_n(\wsb),\ \wsb\in [0,1]^d,$ the tree-based copula density estimate, computed on $\wsb^{(1:n)}$, and given in terms of a binary partition $\Lambda^s_c$ of $[0,1]^d$. The boundaries of sets in the partition $\Lambda^s_c$, and the points in $\wsb^{(1:n)}$ they contain, are mapped back from $[0,1]^d$ to determine a binary partition, $\Lambda^x_f$ say, and a corresponding density estimate in $\Omega$, $\tilde f^{(c)}_n$ say, using the generalised inverse $\widehat{F}^{-1}_i(\wsb),\ i=1,\ldots,d,\ \wsb\in [0,1]^d$.
 	
 	However, this approach has the weakness that the empirical transformation is non-linear and adapted to the data and so, although the $D^*$ condition holds between $\wsb^{(1:n)}$ and $\Lambda^s_c$ in $[0,1]^d$, it does not in general hold between $\xb^{(1:n)}$ and $\Lambda^x_c$ in $\Omega$. When $F$ is skewed, sample points $\wsb^{(1:n)}$ which are evenly distributed over sets in $\Lambda^s_c$ may, on mapping back to $\xb^{(1:n)}$, be concentrated at the boundaries of sets in $\Lambda^x_f$ and may be quite unevenly distributed in those sets. In this case the piecewise constant density estimate $\tilde f^{(c)}_n$ can be a very poor fit to $f$. This is what we observed in experiments, where the resulting HPD set estimate tends to include regions of low probability mass in $F$. In contrast, truncating $\Omega$ down to $\Omega^{(p)}$ and carrying out estimation in $\Omega^{(p)}$ allows for a simple linear mapping between $\Omega^{(p)}$ and $[0,1]^d$ and so the $D^*$ condition holds in both spaces. In our case we get exactly the same partition if we work entirely in $\Omega^{(p)}$, and none of these issues arise.
	 
	\section{Credible set estimation, loss and convergence}\label{sec:credset}
	
	Consider now the case where the target distribution $F$ with density $f$ is a posterior distribution, or some fixed approximation to the posterior. Recall from Condition~\ref{cond3} in Section~\ref{sec:dentree} that $Q^{\alpha}$ is the level-$\alpha/p$ HPD set for the conditional distribution $F^{(p)}$ of $\Xb\sim F$ given $\Xb\in \Omega^{(p)}$. 
	Write $\hat f^{(p)}_n(\xb)=\hat f^{(p)}_n(\xb;\Xb^{(1:n)},\hat\Omega^{(p)}_n)$ for the tree estimate for the density $f^{(p)}$ defined above Eqn.~\ref{eq:Fhatp}. Corollary~\ref{cor6} holds in this case. All properties discussed below hold if $\hat\Omega^{(p)}_n$ is replaced by a fixed set $\Omega^{(p)}$, and the stronger Corollary~\ref{cor4} holds.
	
	\subsection{HPD set estimation} \label{sec:hpdest} We assume the leaf-labels are sorted so that $n_{k}/|\Lambda_k|\ge n_{k+1}/|\Lambda_{k+1}|$ for $k=1,\ldots,K-1$ and take as our estimated HPD set
	\[
	\tilde Q^{\alpha_{n,\tau}}=\bigcup_{k=1}^{\tilde K_\alpha} \Lambda_k
	\]
	where 
	\[
	{\tilde K_\alpha}=\arg\min_{0\le K'\le K} \left|\frac{1}{n}\sum_{k=1}^{K'} n_{k}-\alpha/p\right|,
	\]
	and $\alpha_{n,\tau}=\sum_{k=1}^{\tilde K_\alpha} n_{k}/n$,
	so we include the highest density leaves with combined coverage $\alpha_{n,\tau}=\hat F^{(p)}(\tilde Q^{\alpha_{n,\tau}})$ closest to target. 
	
	\subsection{Alternatives to HPD sets}
	
	When the target distribution is continuous, the HPD set is the set $Q=Q^\alpha$ in $\mathcal{B}$ minimising the volume $\int_{Q} d\xb$ subject to $F(Q)\ge \alpha$. 
	The choice of Lebesgue measure $d\xb$ has the consequence that HPD sets are not reparameterisation invariant: if 
	$r$ is an invertible and differentiable function of $\xb\in \Omega$ then $\Pr(r(\Xb)\in r(Q^\alpha))=\alpha$ but $r(Q^\alpha)$ is not in general the HPD set of $r(\Xb)$ \citep{bernado05}. This motivates \citet{rousseau05} to define a J-HPD set $Q=Q^{\alpha}_J$
	minimising Jeffrey's measure $\int_{Q} \sqrt{|I(\xb)|}d\xb$ subject to $F(Q)\ge \alpha$, with $I(\xb)$ the information matrix. The J-HPD set for $r(\Xb)$ is $r(Q^{\alpha}_J)$ and its coverage is $\alpha$. \citet{druilhetmarin07} show that the J-HPD set can be expressed in the form 
	\[
	Q^{\alpha}_J=\{\xb\in \Omega;f(\xb)/\sqrt{|I(\xb)|}>\gamma_J^\alpha\},
	\]
	with $\gamma_J^\alpha$ chosen to ensure $F(Q^{\alpha}_J)=\alpha$. We can use this to estimate $Q^{\alpha}_J$: proceed as in Section~\ref{sec:hpdest}, but sort the leaves on their $n_k/\sqrt{I_k}|\Lambda_k|$-values in decreasing order, with $I_k=|I(\xb^{(j)})|$ the information evaluated at a sample point $\xb^{(j)}\in \Lambda_k$, and then accumulate leaves till the target coverage is reached. \citet{rousseau05} consider advantages and disadvantages of this class of HPD sets. We have not pursued this further.
	
	\subsection{Loss and loss estimation} \label{sec:loss} Let $\mu(d\xb)=f_\mu(\xb)d\xb$ be a probability distribution defined for sets $Q\in \mathcal B^{(p)}$, absolutely continuous with respect to Lebesgue measure in $\Omega^{(p)}$, with density $f_\mu(\xb)$. We define the loss for estimating set $Q$ when the truth is $Q^{\alpha}$ to be
	\begin{equation}\label{eq:loss}
	    L_\mu(Q,Q^\alpha)=\mu(Q\,\Delta\,Q^\alpha),
	\end{equation}
	the $\mu$-measure of the set difference in Eqn.~\ref{eq:sd}. 
	We use $\mu=F$ in many of our experiments but there can be some advantage in taking $\mu$ more dispersed than $F$ as discussed below.
	
	When the posterior density $f(\xb)\propto \fu(\xb)$ can be evaluated up to a constant in $\xb$, it is possible to estimate $L_\mu(Q,Q^\alpha)$ consistently for any fixed set $Q\in \mathcal{B}^{(p)}$ without needing to know $Q^\alpha$. We use set-difference estimates of the kind introduced in \citet{baillo00} (see that paper for references to earlier related ideas). Our set-estimation algorithm chooses the bandwidth $\tau$ so that no significant improvement in $L_\mu(Q^{\alpha_{n,\tau}},Q^\alpha)$ can be made by varying $\tau$.

	Recall that $\Xb^{(1:n)}$ are $n$ samples distributed according to $F$. 
	For $j=1,\ldots,n$, let $\fu^{(j)}$ denote the $j$'th order statistic of the unnormalised sample density values $\fu(\Xb^{(j)}),\ j=1,\ldots,n$ on the original samples and let 
	\begin{equation}\label{eq:hatgamma}
	\hat\gamma=\fu^{(\lfloor(1-\alpha) n\rfloor)}
	\end{equation} 
	be an estimate of the $(1-\alpha)$-quantile of $\fu(\Xb)$ based on its order statistics.
	For $j'=1,\ldots,m$ let $\Zb^{(j')}\sim \mu$ be $m$ samples which are independent of $\Xb^{(1:n)}$ and distributed according to $\mu$. Let $\Zb^{(1:m)}=\{\Zb^{(1)},\ldots \Zb^{(m)}\}$.
	For sets $Q\in\mathcal{B}^{(p)}$ the estimators 
	 \begin{equation}\label{FP}
	 FP(Q;\Zb^{(1:m)})=\frac{1}{m}\sum_{j=1}^m \mathbbm{1}(\fu(\Zb^{(j)})< \hat\gamma)\mathbbm{1}(\Zb^{(j)}\in Q)
	 \end{equation}
	 and
	 \begin{equation}\label{FN}
	 FN(Q; \Zb^{(1:m)})=\frac{1}{m}\sum_{j=1}^m \mathbbm{1}(\fu(\Zb^{(j)})\ge\hat\gamma)\mathbbm{1}(\Zb^{(j)}\in Q^c),
	 \end{equation}
	 respectively estimate ``false postive'' ($\Zb$ in $Q$ but not in $Q^\alpha$) and ``false negative'' ($\Zb$ not in $Q$ but in $Q^\alpha$) rates on the test set $\Zb^{(1:m)}$. These will not count ``true'' false positives or negatives with respect to $Q^\alpha$ as the threshold $\hat\gamma$ defined in Eqn.~\ref{eq:hatgamma} is only a consistent estimate.  
	 
	 The following Theorem makes explicit our loss-estimation procedure. Results of this kind are well-known in the level-set estimation literature.
	\begin{theorem}\label{thm5}
	 Let $Q\in\mathcal{B}^{(p)}$ be given. If, for sets $A\in\mathcal{B}$, we have $\frac{1}{m}\sum_{j=1}^m \mathbbm{1}(\Zb^{(j)}\in A)\tip \mu(A)$ then the loss estimator
	 \[\hat L_\mu(Q; \Zb^{(1:m)})=FP(Q; \Zb^{(1:m)})+FN(Q; \Zb^{(1:m)}),\]
	 which does not depend on the unknown $Q^{\alpha}$, is a consistent estimator for 
	 \[L_\mu(Q,Q^\alpha)=\mu(Q\setminus Q^\alpha)+\mu({Q^\alpha}\setminus Q)\]
	 in the limit $\min(m,n)\rightarrow\infty$.
	\end{theorem}
	Proof: let $\gamma$ satisfy
	\[
	F(\{\xb\in \Omega: \fu(\xb)>\gamma\})=\alpha,
	\]
	so $\gamma$ is the HPD level set threshold value expressed in terms of the unnormalised density function $\fu$ and $Q^{\alpha}=\{\xb\in \Omega: \fu(\xb)> \gamma\}$. Now $P(\fu(\Xb)>\gamma)=\alpha$ \citep{box65} so $\gamma$ is the $1-\alpha$ quantile of the random variable $\fu(\Xb)$ and $\hat\gamma$ in Eqn.~\ref{eq:hatgamma} is a consistent estimator for $\gamma$. 
    
    We now show that $FN\tip  \mu({Q^\alpha}\setminus Q)$. The proof for $FP\tip \mu(Q\setminus Q^\alpha)$ is similar. First of all, as $\Zb^{(1:m)}$ and $\hat\gamma$ are independent,
	\[
	FN(Q; \Zb^{(1:m)})\tip  \int_{\Omega} \mathbbm{1}(\fu(\xb)> \hat\gamma)\mathbbm{1}(\xb\in Q^c)f_\mu(\xb)\, d\xb
	\]
	as $m\rightarrow\infty$ at every finite $n>0$ by the WLLN. Furthermore,
	\[
	\int_{\Omega} \mathbbm{1}(\fu(\xb)> \hat\gamma)\mathbbm{1}(\xb\in Q^c)f_\mu(\xb)\, d\xb\tip  \int_{\Omega} \mathbbm{1}(\fu(\xb)> \gamma)\mathbbm{1}(\xb\in Q^c)f_\mu(\xb)\, d\xb
	\]
	as $\hat\gamma\tip\gamma$ because the left integral is a continuous function of the level-set threshold $\hat\gamma$. However,
	\begin{align}
	\int_{\Omega} \mathbbm{1}(\fu(\xb)> \gamma)\mathbbm{1}(\xb\in Q^c)f_\mu(\xb)\, d\xb&=\int_{Q^\alpha} \mathbbm{1}(\xb\in Q^c)f_\mu(\xb)\, d\xb\\
	&=\mu({Q^\alpha}\setminus Q),
	\end{align}
	and hence
	\[
	\int_{\Omega} \mathbbm{1}(\fu(\xb)> \hat\gamma)\mathbbm{1}(\xb\in Q^c)f_\mu(\xb)\, d\xb\tip  \mu({Q^\alpha}\setminus Q)
	\]
	as $n\rightarrow\infty$. These limits may be taken in the opposite order, and so $FN\tip  \mu({Q^\alpha}\setminus Q)$ as $\min(m,n)\rightarrow\infty$. [End of proof]
	
	We now discuss the choice of loss-measure $\mu$. For any $\mu$ absolutely continuous with respect to Lebesgue measure in $\Omega^{(p)}$, the loss $L_\mu(Q\,\Delta\, Q^\alpha)$ is minimised when the volume of the error-set $Q\,\Delta\, Q^\alpha$ is equal zero. However, locating the optimal bandwidth $\tau$ is easier if the loss grows rapidly with the volume of $Q\,\Delta\, Q^\alpha$.  If we have a good density estimate $\hat f^{(p)}_n$ then points $\xb\in Q\,\Delta\, Q^\alpha$ will be close to the boundary, $\partial Q^\alpha=\{\xb\in \Omega^{(p)}: q(\xb)=\gamma\}$ say, of the true HPD set, so we expect better choices of $\mu$ to have relatively higher values of $f_\mu(\xb)$ around $\partial Q^\alpha$. This is borne out in our discussion of Figure~\ref{mvn_tau} below. 
	
	We parameterise a family of loss-densities $f_{\mu_\beta}(\xb)\propto q(\xb)^\beta,\ 0\le \beta\le 1$ using tempering. This includes Lebesgue measure and $F$, at $\beta=0$ and $\beta=1$ respectively. We choose $\beta$ to maximise $f_{\mu_\beta}(\xb)$ for $\xb\in\partial Q^\alpha$. The following Lemma summarises this step.
	\begin{lemma}\label{lem:ROT}
	For $\xb\in \Omega^{(p)}$, let $U(\xb)=-log(q(\xb))$ and let $U_0=-log(\gamma)$ give the constant value taken by $U(\xb)$ for $\xb\in\partial Q^\alpha$. 
	Let $\mu_\beta(d\xb)=f_{\mu_\beta}(\xb)d\xb$ with $f_{\mu_\beta}(\xb)\propto \exp(-\beta U(\xb))$ for $\xb\in\Omega^{(p)}$ and let $\Xb_\beta\sim \mu_\beta$. The value, $\beta=\beta^*$ say, maximising $f_{\mu_\beta}(\xb)$ for $\xb\in\partial Q^\alpha$ satisfies
	\[
	E(U(\Xb_{\beta^*}))=U_0.
	\]
	If the distribution of $U(\Xb_{\beta^*})$ is symmetric then $\mu_{\beta^*}(Q^{\alpha})=1/2$.
	\end{lemma}
	The proof is straightforward and is omitted. The last part follows when the median of $U(\Xb_{\beta^*})$ is equal to its average, as $U(\Xb_{\beta^*})<U_0$ if and only if $q(\Xb_{\beta^*})>\gamma$ so $\Xb_{\beta^*}\in Q^{\alpha}$. Lemma~\ref{lem:ROT} gives  
	a handy rule of thumb for choosing $\beta$:
	find a reasonable approximation to $Q^\alpha$ and then choose $\beta$ so that about half the sampled $\Zb$-values fall inside $Q^{\alpha}$.
	
	We experiment with this in Section~\ref{sec:bandwidth_selection} below. We find the curvature of the loss with varying $\tau$ is greater at intermediate $\beta$-values, so the loss is more easily optimised (ie, with smaller test-data sample sizes $m$, since lower precision is needed). However, the advantage of varying $\beta$ is slight. We saw the same in other examples which we do not report. At finite $m$ we have only an estimate of the loss so the location of the minimum can depend on $\beta$, but not asymptotically in $m$, as the exact loss $\mu_{\beta}(Q\, \Delta\, Q^\alpha)$ is minimised by $Q=Q^\alpha$ for every $\beta\ge 0$. The disadvantage over simply using $\mu=F$ is that one must implement a sampler to get test samples $\Zb^{(1:m)}\sim \mu_\beta$. Also, test data $\Zb^{(1:m)}\sim F$ are convenient for estimating the realised coverage $\alpha_{n,\tau}$, which is of independent interest. Unless indicated, we take $\mu=F$ everywhere below.
	
	\subsection{Convergence} \label{sec:converge}
	When the posterior is ``doubly intractible'', it cannot be evaluated up to an overall constant in $\xb$ (recall, this is the parameter, not the data) and we cannot estimate the loss in Theorem~\ref{thm5}. In this case we would like to show convergence of the loss to zero without the need to measure it.
	  We conjecture, but have not proven, that $L_F(\tilde Q^{\alpha_{n,\tau}},Q^\alpha)$ converges to zero in probability with increasing sample size $n$, as the density of $\tilde F^{(p)}_\infty$ introduced at the end of Section~\ref{sec:estimatetruncate} coincides with that of $F$ except on sets of zero measure. 
	  We find in our experiments that our HPD set estimates $\tilde Q^{\alpha_n,\tau}$ do appear to converge and we can at least show that if Condition~\ref{cond:bound} holds, and HPD-set estimates do converge, then they converge to an HPD set. We dont know the coverage of this limit set, but we can estimate it, and drive it towards $\alpha$ in our bandwidth selection in the next section.
	  
	  \begin{condition}\label{cond:bound}
	  The target and approximating densities are bounded in $\Omega$, that is, there exist constants $M,M'<\infty$ satisfying $f(\xb) \le M$ and 
	  $P(\hat f^{(p)}_n(\xb)\le M')\rightarrow 1$ for all $\xb\in\Omega$ (taking $\hat f^{(p)}_n(\xb)=0$ for $\xb\in \Omega\setminus\hat\Omega^{(p)}_n$).
	  \end{condition}
	  
	  \begin{theorem}\label{thm6}
	  Suppose Condition~\ref{cond:bound} is satisfied. If for some $Q^*\in \mathcal{B}^{(p)}$ it holds that 
	  \begin{equation}\label{eq:Fconverge}
	      F(\tilde Q^{\alpha_n,\tau}\,\Delta\,Q^*)\tip 0
	  \end{equation}
	  then $Q^*$ is a level-$F^{(p)}(Q^*)$ HPD set of $F^{(p)}$ (up to sets of $F$-measure zero).
	  \end{theorem}
	  Proof: Take $Q\in \mathcal{B}$ and let $A$ be the set of all points $\xb\in Q, \xb'\in Q^c$ such that $f(\xb)<f(\xb')$. If $A$ is empty then $Q$ is an HPD set. If $A$ has $F$-measure zero then $Q$ differs from an HPD set on a set of measure zero.
	  Take two equal volume hyper-rectangular sets $\Hr,\Hr'\in\mathcal{H}^{(p)}$, satisfying  $\Hr\subset Q^{*}$ and $\Hr'\subset (Q^{*})^c$.
	  If $F(\Hr)\ge F(\Hr')$ for all such pairs then $Q^*$ is an HPD set (up to a set of $F$-measure zero). 
	  
	  As $\Hr'\cap Q^*=\emptyset$ we have $F(\Hr'\cap \tilde Q^{\alpha_n,\tau})\tip 0$ from Eqn.~\ref{eq:Fconverge}. Since $F$ (and $F^{(p)}$) has a density with respect to volume measure, and $f^{(p)}$ is uniformly bounded, we have $|\Hr'\cap \tilde Q^{\alpha_n,\tau}|\tip 0$, and then since $\hat f^{(p)}_n$ is uniformly bounded in probability, $\hat F^{(p)}_n(\Hr'\cap \tilde Q^{\alpha_n,\tau})\tip 0$, and so $\hat F^{(p)}_n(\Hr')-\hat F^{(p)}_n(\Hr'\cap (\tilde Q^{\alpha_n,\tau})^c)\tip 0$. As $\hat F^{(p)}_n(\Hr')\tip F^{(p)}(\Hr')$ by Corollary~\ref{cor6}, we have 
	  \[
	  \hat F^{(p)}_n(\Hr'\cap (\tilde Q^{\alpha_n,\tau})^c)\tip F^{(p)}(\Hr').
	  \] 
	  Similar reasoning leads to 
	  \[
	  \hat F^{(p)}_n(\Hr\cap \tilde Q^{\alpha_n,\tau})\tip F^{(p)}(\Hr).
	  \]
	  By the HPD-estimate construction rule in Section~\ref{sec:hpdest}, $\hat f^{(p)}_n(\xb)\ge \hat f^{(p)}_n(\xb')$ for $\xb\in \tilde Q^{\alpha_n,\tau}$ and $\xb'\in (\tilde Q^{\alpha_n,\tau})^c$.
	  The set volumes satisfy $|\Hr\cap \tilde Q^{\alpha_n,\tau}|\tip |\Hr|$ and $|\Hr'\cap (\tilde Q^{\alpha_n,\tau})^c|\tip |\Hr'|$ with $|\Hr|=|\Hr'|$, so for some $\epsilon\ge 0$,
	  \[
	  \hat F^{(p)}_n(\Hr\cap \tilde Q^{\alpha_n,\tau})-\hat F^{(p)}_n(\Hr'\cap (\tilde Q^{\alpha_n,\tau})^c)\tip \epsilon,
	  \]
	  and hence $F^{(p)}(\Hr)\ge F^{(p)}(\Hr')$. [End of proof]
	  
	When the posterior is doubly intractable, we cannot estimate the full loss. 
    By Lemma~\ref{lem1} below, the loss is always lower bounded by $|\alpha_{n,\tau}-\alpha|$, with the bound achieved if $Q^{\alpha_{n,\tau}}$ is an HPD set. However, from Theorem~\ref{thm6} the estimator ${\tilde Q}^{\alpha_{n,\tau}}$ converges to an HPD set (if it converges) so in that case the loss is close to the lower bound at large $n$. In the next section we minimise this lower bound on the loss when we cant evaluate the loss. The bound is easily estimated using test samples.
	\begin{lemma}\label{lem1} 
 	 Let $\mathcal{C}_{\widetilde \alpha}=\{Q\in \mathcal{B}: F(Q)=\widetilde \alpha\}$ be the set of all sets with fixed coverage $\widetilde \alpha$, so that HPD set $Q^{\tilde\alpha}\in \mathcal{C}_{\widetilde \alpha}$. For all $Q\in \mathcal{C}_{\widetilde \alpha}$ we have $L_F(Q,Q^{\alpha})\ge L_F(Q^{\widetilde{\alpha}},Q^\alpha)$ with \[L_F(Q^{\widetilde{\alpha}},Q^\alpha)=|\widetilde\alpha-\alpha|.\]
 	\end{lemma}
 	Proof: the loss at $Q$ can be written
 	\[
    L_F(Q,Q^\alpha)=F(Q)+F(Q^\alpha)-2F({Q}\cap Q^\alpha)
    \]
    and so $L_F({Q},Q^{\alpha})=\widetilde\alpha+\alpha-2F({Q}\cap Q^\alpha)$.
 	If $\widetilde{\alpha}<\alpha$ then $F({Q}\cap Q^\alpha)$ is maximised over $Q\in \mathcal{C}_{\widetilde \alpha}$ by any set of $F$-measure $\widetilde \alpha$ which is a subset of $Q^\alpha$. However HPD sets are nested by coverage, so $Q^{\widetilde{\alpha}}$ is one such set. In this case $L_F({Q}^{\widetilde\alpha},Q^\alpha)=\alpha-\widetilde\alpha$.
 	If $\widetilde{\alpha}>\alpha$ then $Q^{\widetilde{\alpha}}$ contains $Q^{{\alpha}}$ as a subset, minimising $L$ at $L_F({Q}^{\widetilde\alpha},Q^\alpha)=\widetilde\alpha-\alpha$. [End of proof]
	
	\subsection{Bandwidth selection}\label{sec:bandwidth_selection}

    In this section we explain how the ``bandwidth'' or smoothing parameter $\tau$ appearing in DSP Algorithm~\ref{alg:detcore} is selected and give two algorithms for estimating HPD sets covering the cases where we can (Algorithm~\ref{alg:tractablepost}) or cannot (Algorithm~\ref{alg:DI}) estimate the full loss, $L_F({\tilde Q}^{\alpha_{n,\tau}}, Q^\alpha)$ in Eqn.~\ref{eq:loss} using the estimator in Theorem~\ref{thm5}. 
    Simulation of training and test samples, respectively  $\Xb^{(1:n)}$ and $\Zb^{(1:m)}$, is relatively expensive so we assume the total number of samples $n+m$ is fixed. Tree construction is relatively rapid so we search over a grid of $\tau$ values at fixed $n$ using the test samples
    to estimate the loss, either $L_F({\tilde Q}^{\alpha_{n,\tau}}, Q^\alpha)$ or $|\widetilde\alpha_{n,\tau}-\alpha|$.
    
\begin{figure}[!h] 
		\centering
		\includegraphics[scale=.3]{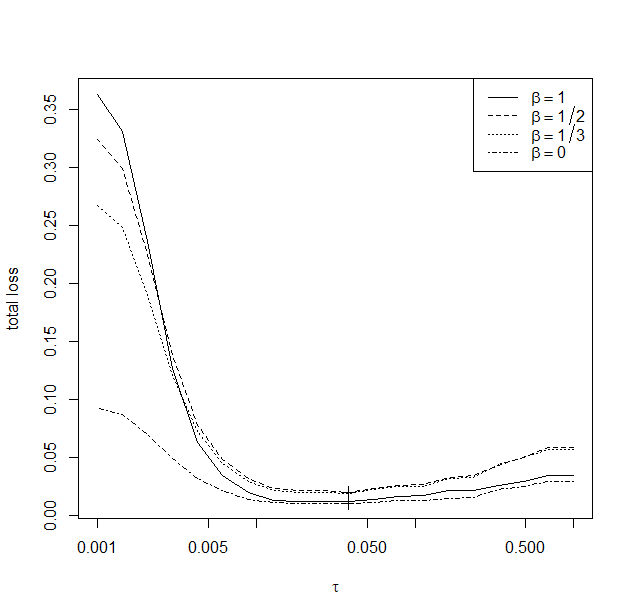} 
		\includegraphics[scale=.3]{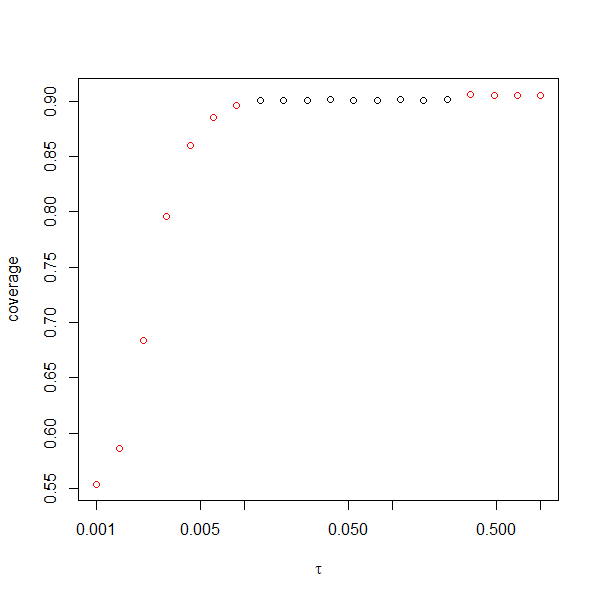} 
		\includegraphics[scale=.3]{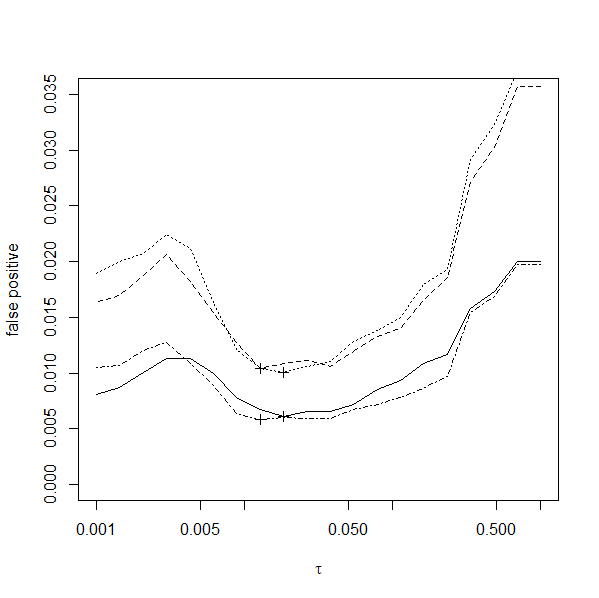} 
		\includegraphics[scale=.3]{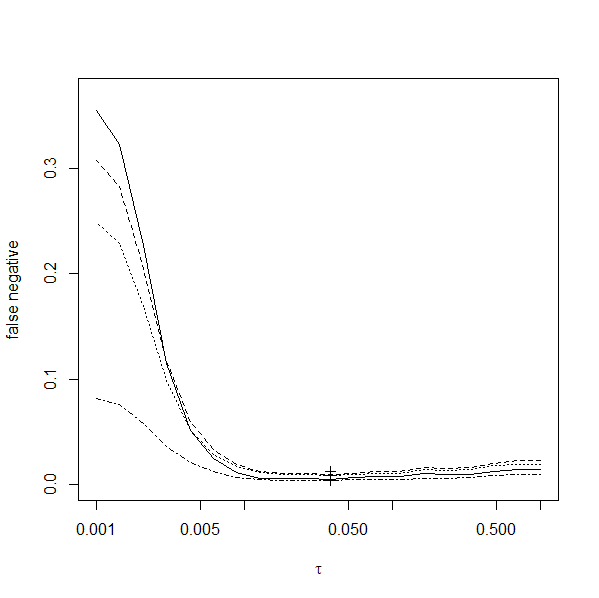} 
		\caption{HPD set estimation summary for a bivariate Gaussian distribution: $\tilde{Q}^{\alpha_{n,\tau}}$ is estimated at each $\tau$-value on the $x$-axes; different measures are shown on the $y$-axes; total loss estimate $\hat L_{\mu_\beta}$ at inverse temperature $\beta=0,1/3,1/2,1$ (top-left), coverage estimate $\hat\alpha_{n,\tau}$ colored red(fail)/black(pass) by hypothesis test $H_0: \alpha_{n,\tau}=\alpha$ outcome (top-right), false positive rate (bottom-left, Eqn.~\ref{FP}), and false negative rate (bottom-right, Eqn.~\ref{FN}). Minimum rates among points passing coverage test top-right marked by cross bottom row.}\label{mvn_tau}
	\end{figure}

    Figure \ref{mvn_tau} illustrates how the HPD set estimation rule given in Section~\ref{sec:hpdest} and the loss in Section~\ref{sec:loss} behaves with $\tau$ on a bivariate Gaussian target distribution for one sample set $\xb^{(1:n)}\sim N([0,0],I_{2\times 2})$ and one test set with loss-measure $\zb^{(1:m)}\sim N([0,0],\beta^{-1}I_{2\times 2})$ for sample sizes $n=3\times 10^5$ and $m=3\times 10^4$. The exact HPD set is $Q^\alpha=\{\xb\in \mathbb{R}^2: |\xb|^2<-2\log(1-\alpha)\}$. The estimated loss $\hat L_{\mu_\beta}$ for $\tilde{Q}^{\alpha_{n,\tau}}$ (top left) has a classical shape corresponding to under-smoothing (at small $\tau$) and over-smoothing (large $\tau$). The temperature $\beta^{-1}$ in the loss-measure $\mu_\beta(d\xb)=N(d\xb;[0,0],\beta^{-1}I_{2\times 2})$ is equal to the variance of the test samples with $\mu_\beta$ proportional to Lebesgue measure at $\beta=0$. The value of $\beta$ maximising the loss-measure density $f_{\mu_\beta}$ on the boundary of the true HPD set $\partial Q^\alpha$, which we can calculate here and is at $|\xb|^2=-2\log(1-\alpha)$, in Lemma~\ref{lem:ROT} is $\beta^*=-1/\log(1-\alpha)$, so that $\beta^*\simeq 0.43$. The loss function top left in Figure~\ref{mvn_tau} at $\beta=1/3$ and $\beta=1/2$ is slightly more cup-shaped than is the case for the uniform ($\beta=0$) or target ($\beta=1$) distributions, but the advantage is slight. The minimum for all $\beta$ has the same location, as expected. The False Positive ($FP$, lower left) and False Negative (lower right) counts have minima at similar values of $\tau$. This is because the coverage ($\alpha_{n,\tau}$, upper right) is flat for large $\tau$ at values close to $\alpha$. In this regime $\alpha_{n,\tau}\simeq\alpha$ is effectively fixed.
    Then since
    \[L_F(\tilde{Q}^{\alpha_{n,\tau}},Q^\alpha)=\alpha-\alpha_{n,\tau}+2F(\tilde{Q}^{\alpha_{n,\tau}} \backslash Q^\alpha)\]
    the loss is minimised when we minimise $F(\tilde{Q}^{\alpha_{n,\tau}} \backslash Q^\alpha)$, which is estimated by the False Positives alone. Similar reasoning applies to the False Negatives. Our algorithm exploits this behavior. 
    Let $\tau_1>\tau_2>\ldots >\tau_J$ be a sequence of $J$ smoothing parameters and  for $j=1,\ldots,J$ let 
    \begin{equation}\label{eq:hatalpha}
    \hat\alpha_{n,\tau_j}=m^{-1}\sum_i \mathbbm1(Z^{(i)}\in \tilde Q^{\alpha_{{n,\tau_j}}})    
    \end{equation} be our estimate of $\alpha_{n,\tau_j}$ on the test samples. For $j=1,\ldots,J$, the sample size $m$ determines a search resolution through the CLT approximation 
    \[
    \hat\alpha_{n,\tau_j}\sim N(\alpha_{n,\tau_j},\alpha_{n,\tau_j}(1-\alpha_{n,\tau_j})/m),
    \] 
    with $m$ replaced by $m/t_\alpha$ for MCMC output with integrated auto-correlation time $t_\alpha$. In Algorithm~\ref{alg:DI}, where the loss is approximated by the lower bound $|\widetilde\alpha_{n,\tau}-\alpha|$, we carry out a size-$\delta$ hypothesis test with $\delta=0.05$, null $H_0: \alpha_{n,\tau_j}=\alpha$ and return the HPD set estimate $\tilde Q^{\alpha_{{n,\tau_j}}}$ from the smallest $\tau_j$-value at which the test passes. Our choice of the smallest $\tau$-value reflects the intuition that smaller bandwidth at fixed coverage gives better adapted sets, and is born out in experiment. In Algorithm~\ref{alg:tractablepost}, which applies when the loss can be estimated, we take the $\tau_j$-value yielding the least false-positive count among those passing the test. The power, $B$ say, of the test is the probability to reject the null when $|\alpha_{n,\tau}-\alpha|$ is as large as $\epsilon$, where $\epsilon$ is a user-specified tolerance (we take $\epsilon\simeq 0.03$ below). With an ESS of order $m=3\times 10^4$, the power is close to one at $\alpha=0.5$ (worst case). At $\alpha=0.9$ say the same power is achieved at much smaller values of the tolerance. This is the value of $m$ used in Section~\ref{sec:sim}.

    \begin{algorithm}[t]
	\caption{$\mathstrut$ Level-$\alpha$ HPD set estimation for intractable density $f$}
	Input: target coverage $\alpha$, truncation-level $p$, $J$ values for the smoothing parameter $\tau_1>\tau_2>...>\tau_J$, training and test datasets ($\Xb^{(1:n)}$ and $\Zb^{(1:m)}$) and test size $\delta$.
	
	Output: $\widetilde{Q}^{\alpha_{n,\tau^*}}$.  
	
	\begin{algorithmic}[1]\label{alg:DI}
	
	\STATE Compute the estimated truncation region $\hat\Omega^{(p)}$ 
	\FOR {$j=1,...,J$}
		\STATE Estimate $\hat{f}^{(p)}$ for $\tau_j$ using Algorithm~\ref{alg:detcore} with inputs $\Xb^{(1:n)}$ and $\hat\Omega^{(p)}$.
		\STATE Compute $\widetilde{Q}^{\alpha_{n,\tau_j}}$ as in Section~\ref{sec:hpdest} and estimated coverage $\hat\alpha_{n,\tau_j}$ using $\Zb^{(1:m)}$.
		\STATE Test $H_0: \alpha_{n,\tau_j}=\alpha$ with size $\delta$. 
	\ENDFOR
	\STATE Among the $\tau$ values passing the hypothesis test, choose the smallest value $\tau^*$.
	\end{algorithmic}
\end{algorithm}

    \begin{algorithm}[t]
	\caption{$\mathstrut$ Level-$\alpha$ HPD set estimation for tractable density $f\propto q$}
	Input: target coverage $\alpha$, truncation-level $p$, $J$ values for the smoothing parameter $\tau_1>\tau_2>...>\tau_J$, training and test datasets ($\Xb^{(1:n)}$ and $\Zb^{(1:m)}$), test size $\delta$ and sorted relative density values $q^{(1)}<q^{(2)}<\ldots,q^{(m)}$ evaluated on $Z^{(1:m)}$.
	
	Output: $\widetilde{Q}^{\alpha_{n,\tau^*}}$.  
	
	\begin{algorithmic}[1] \label{alg:tractablepost}
	
	\STATE Compute the estimated truncation region $\hat\Omega^{(p)}$ 
	\FOR {$j=1,...,J$}
        \STATE Estimate $\hat{f}^{(p)}$ for $\tau_j$ using Algorithm~\ref{alg:detcore} with inputs $\Xb^{(1:n)}$ and $\hat\Omega^{(p)}$.
		\STATE Compute $\widetilde{Q}^{\alpha_{n,\tau_j}}$ as in Section~\ref{sec:hpdest} and estimated coverage $\hat\alpha_{n,\tau_j}$ using $\Zb^{(1:m)}$.
		\STATE Test $H_0: \alpha_{n,\tau_j}=\alpha$ with size $\delta$ and compute the false positive rate $FP(\tilde{Q}^{\alpha_{n,\tau_j}},\Zb^{(1:m)})$ using the input $q^{(j)},\ j=1,\ldots,m$. 
	\ENDFOR 
	\STATE Among the $\tau$ values passing the hypothesis test, choose the value $\tau^*$ minimizing the false positive rate.
	\end{algorithmic}
\end{algorithm}


	
	It is possible that the test may reject at all $\tau$. This never occurred in our examples but might when the power is very high and we detect departures from $\alpha$ much smaller than our tolerance $\epsilon=0.03$. However it may indicate $n$ is too small and more samples are needed. If this had occurred we would lower $m$ for the purpose of testing, decreasing the resolution, but use the full test set to estimate the loss, so we had an accurate measure of the quality of a potentially inaccurate set estimate. Also, in practice in our setting the value of $p$ defining the truncation set $\Omega^{(p)}$ is so close to one that in the examples below we target a coverage $\alpha$ rather than $\alpha/p$ as these are not distinguished at the precision of $\hat\alpha_{n,\tau^*}$.
	
\section{Calibrating HPD sets for approximate posteriors}\label{sec:calibtheory}

Suppose we have an HPD set estimate $\tilde Q^{\alpha_{n,\tau}}$ and we wish to calibrate it, that is, we wish to estimate $\alpha_{n,\tau}$. Up to this point we have assumed test samples $\Zb\sim F$ are available. In this case calibration is straightforward using $\hat\alpha_{n,\tau}$ in Eqn.~\ref{eq:hatalpha}.
We now consider calibration in the case where an HPD estimate has been formed using samples from a distribution $F$ which only approximates the distribution of real interest, and while samples from $F$ are available, samples from the real target distribution are not. This is common in Bayesian inference. In this section we explain briefly how calibration can be done, following \citet{Lee2019} and \citet{Xing2019}. We give example applications
in Section~\ref{sec:calibexample}.

Let $\pi(\bth),\ \bth\in \Omega$ be a prior density on parameter space $\Omega$. If data $\Yb\in\mathcal{Y}$ have observation model $\Yb\sim p(\cdot|\xb)$ and data $\Yb=\yobs$ are measured then the exact posterior is
\[
\pi(\bth|\yobs)\propto p(\yobs|\bth)\pi(\bth).
\]
Let $\bTh\sim \pi(\cdot),\ \bTh\in\Omega$ be a multivariate random variable with density equal to the prior.
Denote by $\tilde\pi(\xb|\yobs)$ the approximate posterior density on $\Omega$ and take the HPD-target density $f$ in Section~\ref{sec:dentree} to be $f(\xb)=\tilde\pi(\xb|\yobs),\ \xb\in\Omega$.
The approximating density $\tilde\pi(\xb|\yobs)$ may be defined explicitly, as in variational inference, or implicitly, as in ABC. In the first case $\tilde Q^{\alpha_{n,\tau}}$ can estimated using Algorithm~\ref{alg:tractablepost} and in the second case Algorithm~\ref{alg:DI} must be used. In either case $\tilde Q^{\alpha_{n,\tau}}$ is an approximate HPD set for an approximate density. We wish to estimate $c(\yobs)$ where
\[
c(\yb)=P(\bTh\in \tilde Q^{\alpha_{n,\tau}}|\Yb=\yb),\quad \yb\in\mathcal{Y}.
\]
We may alternatively write $c(\yobs)=\pi(\tilde Q^{\alpha_{n,\tau}}|\yobs)$, the probability mass the exact posterior puts on $\tilde Q^{\alpha_{n,\tau}}$. Notice that $c(\yobs)$ is not in general equal to $\alpha_{n,\tau}$ since $\alpha_{n,\tau}=\tilde \pi(\tilde Q^{\alpha_{n,\tau}}|\yobs)$, the probability the approximate posterior puts on $\tilde Q^{\alpha_{n,\tau}}$.

Estimation of $c(\yobs)$ cannot be straightforward, as we cannot even sample the exact posterior $\pi(\bth|\yobs)$. However, \cite{Xing2019} give a regression based estimator using ideas related to ABC, but with the simpler goal of estimating a certain posterior expectation and not sampling the posterior. For $j=1,\ldots,n^*$ let $\bTh^{(j)}=\bth^{(j)}$ with $\bTh^{(j)}\sim\pi(\cdot)$ and $\Yb^{(j)}=\yb^{(j)}$ with $\Yb^{(j)}\sim p(\cdot|\bth^{(j)})$
be $n^*$ samples from the true generative model $\pi(\bth)p(\yb|\bth)$ and let 
\[
C^{(j)}=\mathbbm{1}(\bTh^{(j)}\in \tilde Q^{\alpha_{n,\tau}}).
\]
Since $\bTh^{(j)}|\Yb^{(j)}=\yb^{(j)}$ has density $\pi(\cdot|\yb^{(j)})$, we have
\[
C^{(j)}|\Yb^{(j)}=\yb\ \sim\ \mbox{Bernoulli}(c(\yb)),\quad \yb\in\mathcal{Y}.
\]
If for $j=1,\ldots,n^*$ the procedure realises $C^{(j)}=c^{(j)}$ then we can regress the ``data'' $c^{(j)}$ on the ``covariates'' $\yb^{(j)}$ over $\mathcal{Y}$ to obtain an estimate $\hat c(\yb)$ of the calibration function $c(\yb),\ \yb\in\mathcal{Y}$ over data space. This is logistic regression and the calibration estimator is $\hat c(\yobs)$. We carry out the regression using the R package BART \citep{chipman10}. We might fit a Generalised Additive Model. However, the regression is over data space and \citet{Xing2019} find that BART provides more robust estimates on relatively higher dimensional data spaces. In our examples the dimension of data space is low, as our examples have sufficient statistics.

\section{Experiments}\label{sec:sim}
	
	In this section our HPD-set estimates are examined numerically and compared to existing methods (BGHM and SR below) and some natural alternatives of our own construction (KDE and Cluster below). In Section~\ref{sec:choosemn} we check how coverage depends on the sample size $n$. The choice of $n$ will depend on the dimension of $\Omega$ and the nature of the target credible set and would in practice be decided by the available run-time. We find algorithms~\ref{alg:DI} and \ref{alg:tractablepost} can be applied on sample sizes up to $O(10^6)$ in our examples. In section~\ref{sec:comparemethods} existing nonparametric density estimators are compared on simple synthetic examples with known properties. In Section~\ref{sec:real} credible set estimation is illustrated on some real examples. One point here is that the dimension doesn't have to be large for joint HPD sets to be interesting. In Section~\ref{sec:calibexample} we illustrate the calibration methods of Section~\ref{sec:calibtheory} for ABC-approximate credible sets. We also measure the coverage of credible sets computed using a variational approximation. 
	
	The following approximations were used in our implementation. When we construct the tree in Algorithm~\ref{alg:detcore} we calculate $D^*$ in Eqn.~\ref{eq:dstar} taking the supremum over a lattice of splits only (ie, the DSP algorithm of \citet{Li2016}). We use an estimated truncation set $\hat\Omega^{(p)}$ taken to be the smallest hyper-rectangular set containing all the sample points, so that $p=1-2/n$. We ignore this truncation in setting the target level (ie we use $\alpha$ rather than $\alpha/p$). 
	However, we have a loss-estimate (when we use Algorithm~\ref{alg:tractablepost}) and a calibration procedure (when we use Algorithm~\ref{alg:DI}) and so we can measure the quality of any set we produce. In a finite-sample setting this would be necessary anyway, and provides assurance that we are producing useful HPD set estimates.
	All studies are implemented using the processor, Intel(R) Xeon(R) Gold 6142 CPU 2.60GHz. 
	
	\subsection{Choice for $n$ and $m$}\label{sec:choosemn}
	
\begin{figure}[!h] 
		\centering
		\includegraphics[scale=.3]{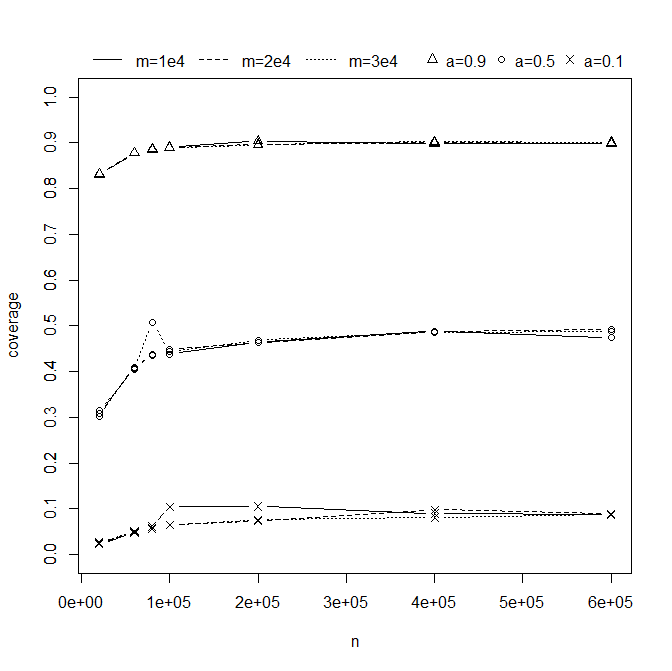}  \includegraphics[scale=.3]{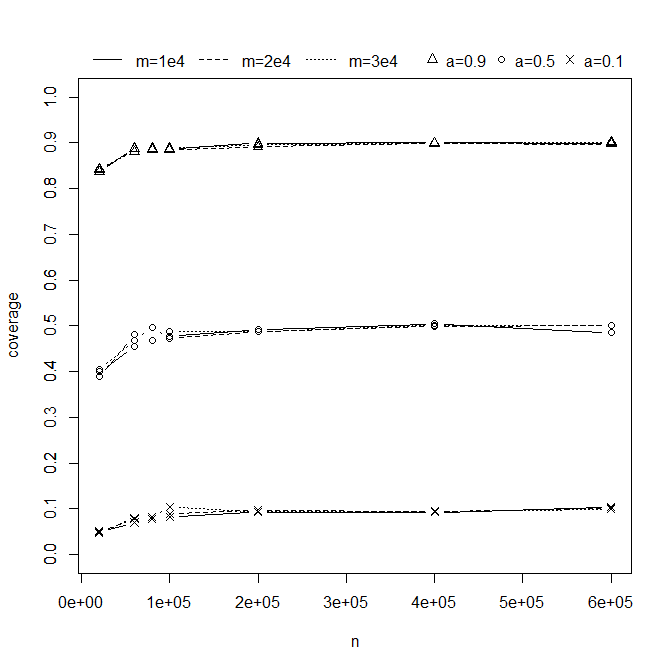}
		\caption{Coverage of HPD set estimation for a bivariate (Left) and a 10-dimensional (Right) Gaussian distributions for a range of $n$ and $m$ values.}\label{mvn_tree}
	\end{figure}

	In this section we investigate the effect of different choices of $n$ on the quality of HPD-set estimates, focusing on coverage of $\tilde{Q}^{\alpha_{n,\tau}}$. We consider two standard multivariate Gaussian distributions $N\left([0,..,0], I_{p\times p} \right)$ with $p=2$ and $p=10$ respectively. We estimate the HPD sets $\tilde{Q}^{\alpha_{n,\tau}}$ and their coverage $\hat\alpha_{n,\tau}$ and plot coverage against $n$ in Figure~\ref{mvn_tree}. At the $m$-values we consider (top of left and right panels) the precision is sufficient to see the $n$ dependence. Comparing the left and right panels, convergence seems to be insensitive to dimension and target $\alpha$ at these relatively low dimension values and large $n$ values, approaching the nominal value as $n$ increases and within measurement precision (not shown) by $n\geq 2\times 10^5$. We expect this sort of behavior to generalise to other distributions at this range of dimension values, unless the HPD sets are very fragmented or present particularly challenging structure. In the examples we take $n=3\times 10^5$. However, as those examples are based on MCMC output, the actual ESS of independent samples was often much lower. In this setting the coverage and loss estimates coming from the test samples play an important role.
	
	\subsection{HPD set estimation comparison}\label{sec:comparemethods}
	
    In this section we compare six HPD set estimators for a range of constructed targets. We are interested in estimation quality (as measured by coverage and loss) and run-time. The runtime comparisons in this section are based on a simple implementation of Algorithm~\ref{alg:detcore} which does not exploit the speed-ups suggested in \cite{Li2016}. They do at least show that the relatively large sample sizes we consider ($n=3\times 10^5$) are easily within range for practical applications.
    
	\begin{enumerate}
	    \item[(a)] KDE - HPD sets from Kernel Density Estimates using the R-package \texttt{ks} in which it is feasible for up to six dimensional densities. If $f^{KDE}$ is the estimated density and $q_{KDE}^{(j)},\ j=1,\ldots,n$ are the order statistics of $f^{KDE}$ evaluated at the training data, $f^{KDE}(\Xb^{(1:n)})$  (see Section~\ref{sec:loss}) then $\hat\gamma^{KDE}=f^{KDE}(\Xb^{(j^*)})$ where the index $j^*$ gives the training sample $\Xb^{(j^*)}$ that falls at the threshold density, $f^{KDE}(\Xb^{(j^*)})=q_{KDE}^{(\lfloor(1-\alpha)n\rfloor)}$. The HPD set is $Q^{KDE}=\{\xb\in\Omega: f^{KDE}(\xb)>\hat\gamma^{KDE}\}$.
	    This is an $O(n^2d)$ method for our normal kernels. 
	    This yields an HPD set $Q^{KDE}$ for $f^{KDE}$ which is not a coverage $\alpha$ set for $f^{KDE}$ but does contain a fraction $\alpha$ of the training samples $X^{(1:n)}$. We find this yields a better approximation and so favors KDE.
	    \item[(b)] Clustering - HPD sets from a Gaussian mixture model with $J_{Clust}$ components fitted using the R-package \texttt{mclust}. Mixture parameters are estimated using an EM algorithm with covariance structure and number of clusters (up to a maximum of $J_{Clust}=10$ here) are chosen using the BIC. If $f^{Clust}$ is the estimated density then the HPD set estimate $Q^{Clust}$ is computed as for $Q^{KDE}$ with $f^{KDE}$ replaced by $f^{Clust}$ in (a). This is $O(nd)$ per EM step. The number of iterations cannot be quantified in general and EM may not converge to the MLE for the training data $\Xb^{(1:n)}$. The memory required to represent the credible set is $O(J_{Clust}d^2)$, typically small.
	    \item[(c)] BGHM \citep{besag1995} - Product of univariate marginal equal tail probability intervals on each dimension with each interval-coverage targeting  $\alpha_{BGHM}\ge\alpha$. The value of $\alpha_{BGHM}$ is determined online by the requirement that the coverage of the joint credible set given by the product of univariate intervals targets $\alpha$.
	    \item[(d)] SR \citep{sorbye2011} - as BGHM with equal-tail marginal intervals replaced by univariate marginal HPD sets. The level $\alpha_{SR}\ge\alpha$ is adjusted so that the product of HPD-sets has coverage targeting $\alpha$. 
	    \item[(e)] DET1 - our Algorithm~\ref{alg:DI} minimising the lower bound for the loss in Lemma~\ref{lem1} with $J_{DET}=10$ $\tau$-values on $[0.01,0.5]$. In our implementation a truncation-set $\hat\Omega^{(p)}$ was estimated (even when not needed) using the BGHM method (c) with $p=1-2/n$. 
	    Runtime complexity is $O(nd)$ at best and $O(n^2d)$ at worst with $O(n\log(n)d)$ if the tree is balanced. However, the actual computations needed are mainly inequalities and the experiments in \citet{Li2016} provide a better guide to real relative cost against KDE. The memory required to represent the credible set in DET1 and DET2 is $O(Kd)$ for $K$ leaves. This is potentially as large as $O(nd)$.
	    \item[(f)] DET2 - Algorithm~\ref{alg:tractablepost}. This minimises the loss in Theorem~\ref{thm5}; otherwise as DET1 in (e). 
	\end{enumerate}

The estimators KDE and Clustering are defined by level-set thresholds.
The point-in-set query (evaluation of $\mathbbm{1}(\xb\in Q^{KDE})$ for example) is handled by evaluating $\mathbbm{1}(f^{KDE}(\xb)>\hat\gamma^{KDE})$. However, there is no explicit geometric representation of the set and this limits its usefulness. For example, we have no immediate access to set topology. The point-in-set query might just as well be handled by simply evaluating $\mathbbm{1}(q(\xb)>\hat\gamma)$ with $\hat\gamma$ given in Section~\ref{sec:loss} without the need to estimate a density at all.
We suggest them as natural alternatives for comparison purposes and not as clearly useful set-estimators.
The DET2 estimator uses a level-set threshold $\hat\gamma$ to evaluate loss but then (like DET1) yields a binary tree-representation of the set. The point-in-set query is evaluated using the inequalities that define the tree.
The BGHM, SR estimators give a geometric representation of the HPD set without density evaluation (or estimation).

We now give four test distributions.

\begin{figure}[!h] 
	\centering
	    \includegraphics[width=7cm]{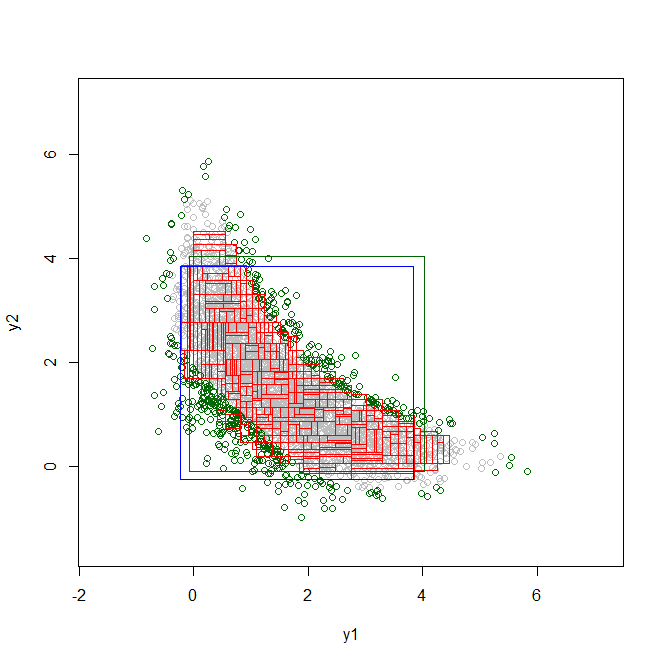}
	    \includegraphics[width=7cm]{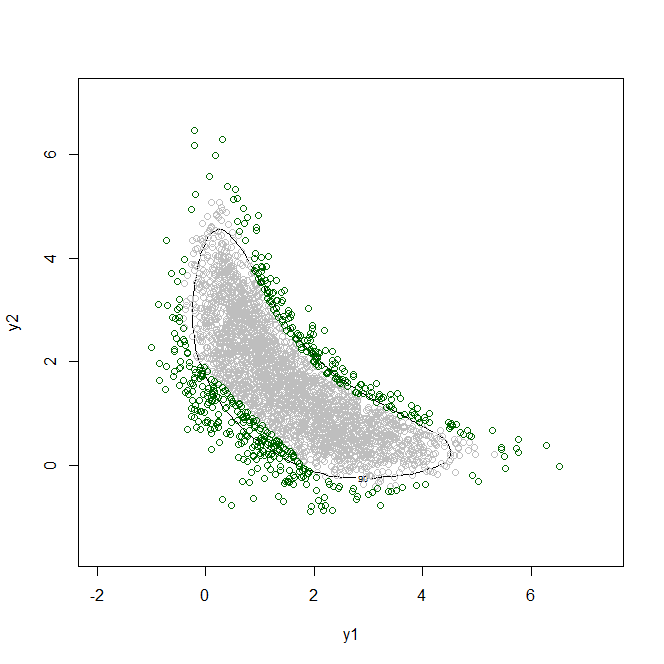}
	    \includegraphics[width=7cm]{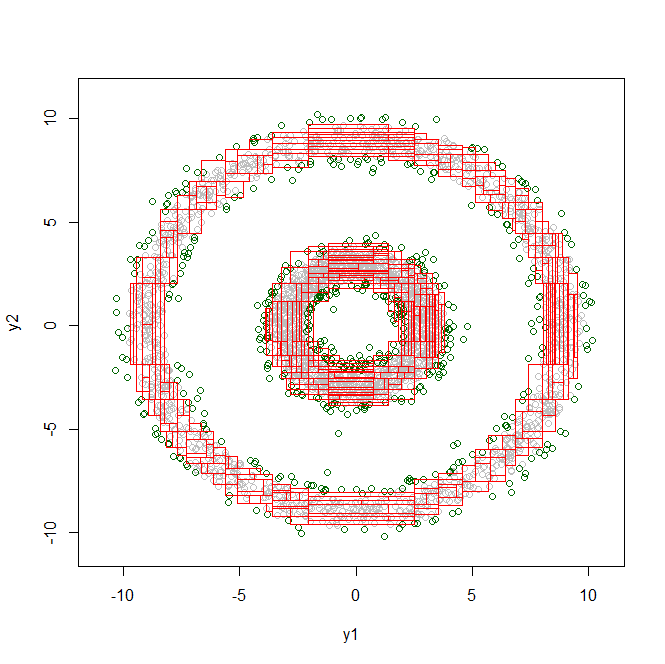}
	    \includegraphics[width=7cm]{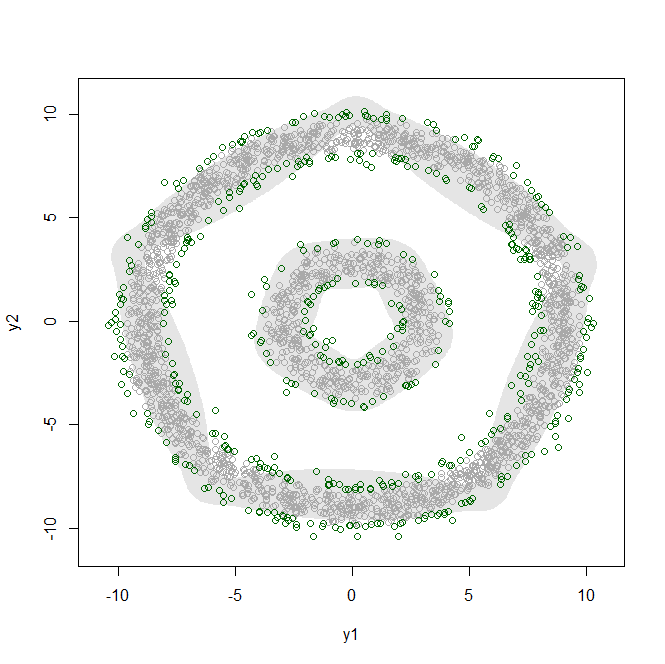}
		\caption{5000 samples from the banana shaped distribution (top row, i) and a donut shaped distribution (bottom row, ii); Points in and outside of $Q^\alpha$ are marked by grey and dark-green respectively. HPD set estimates using DET2 (red boxes, top/bottom-left plots), KDE (black curve, top-right plot), SR (blue box, top-left plot), BGHM (dark green box, top-left plot) and Clustering (light grey area, bottom-right plot). 
		}\label{fig.hpd}
	\end{figure}

	\begin{enumerate}
		\item[(i)] Banana shaped distribution (see Figure~\ref{fig.hpd}, top) for $\xb\in \mathbb{R}^2$ analysed and sampled using the R-package \texttt{bayesm} with parameter values $A=0.5, B=0, C1=3, C2=3$.
		\item[(ii)] Concentric donuts (see Figure~\ref{fig.hpd}, bottom) shaped distribution for $\xb =(r\cos\theta,r\sin\theta)$ where $\theta\sim U[0,2\pi]$ and $r\sim 0.5N(3,0.5^2)+0.5N(9,0.5^2)$. 
		\item[(iii)] Skewed normal distribution for $\xb\in \mathbb{R}^{10}$, $f(\xb)=2N(\mu,\Sigma)\Phi(\beta \xb)$ with $\mu=(0,...,0)$, $\Sigma=I_{10\times 10}$ and $\beta=[-5,...,-1,1,...,5]$ analysed and sampled using the R-package \texttt{EMMIXskew}.
		\item[(iv)] Posterior distribution $\pi(\pmb\mu,\Sigma|\pmb y_1,...,\pmb y_{100})$ where $\pmb y_i\sim N(\pmb\mu,\Sigma)$, $i=1,...,100$. The true parameters are $\pmb\mu=(0,0,0,0)$, $\Sigma_{i,i}=6$ and $\Sigma_{i,j}=5$, $\forall i,j=1,...,4$. The priors are $\pmb \mu \sim N(0,100 I_{4\times 4})$ and, $\Sigma\sim InvWishart(I_{4\times 4},6)$. This makes a 14 dimensional posterior distribution; 4 for $\pmb \mu$ and 10 for $\Sigma$. 
		\end{enumerate}
   
\begin{figure}[!h] 
   	\centering
   	\includegraphics[width=17cm,height=5cm]{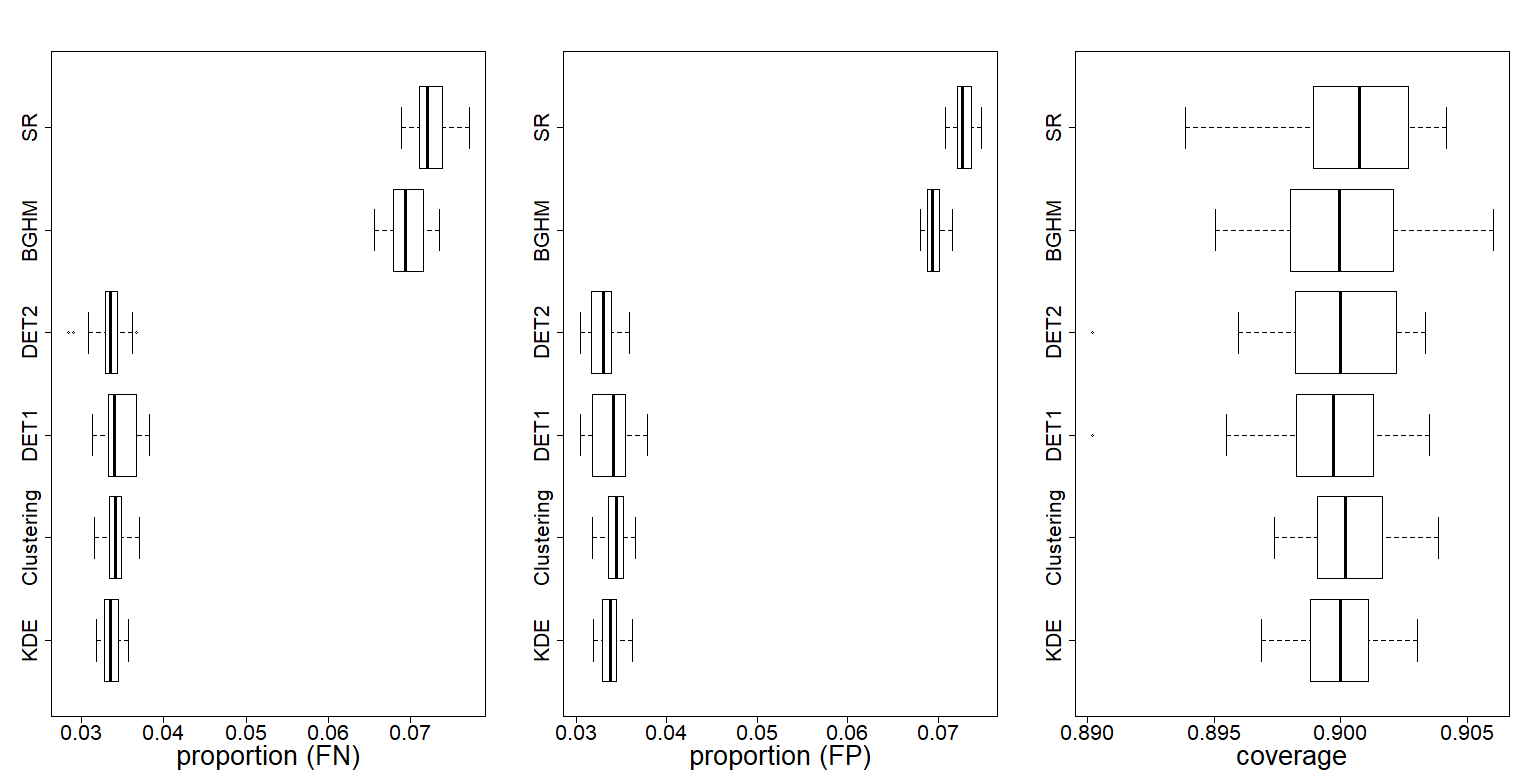}
   	\includegraphics[width=17cm,height=5cm]{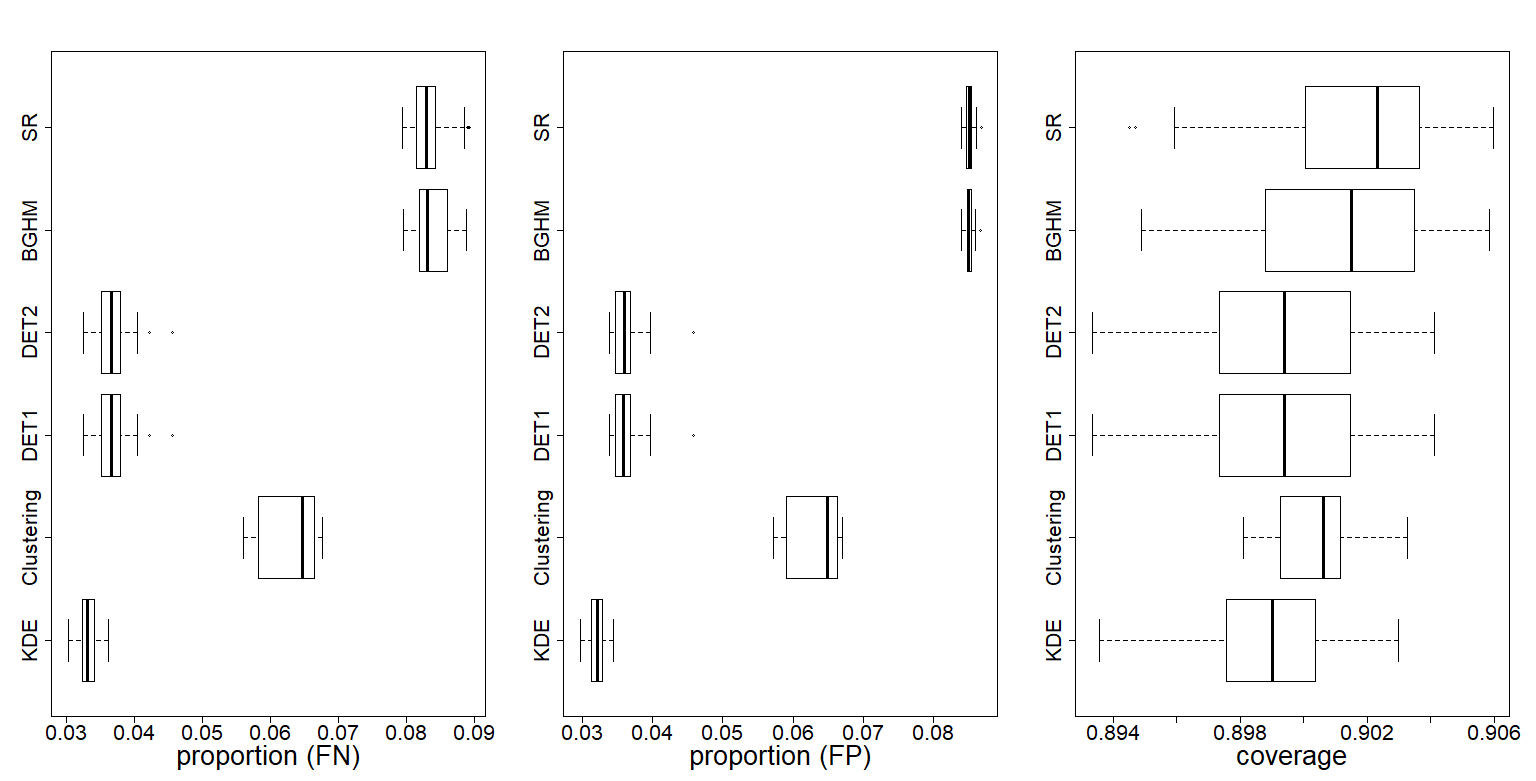}
   	\includegraphics[width=17cm,height=5cm]{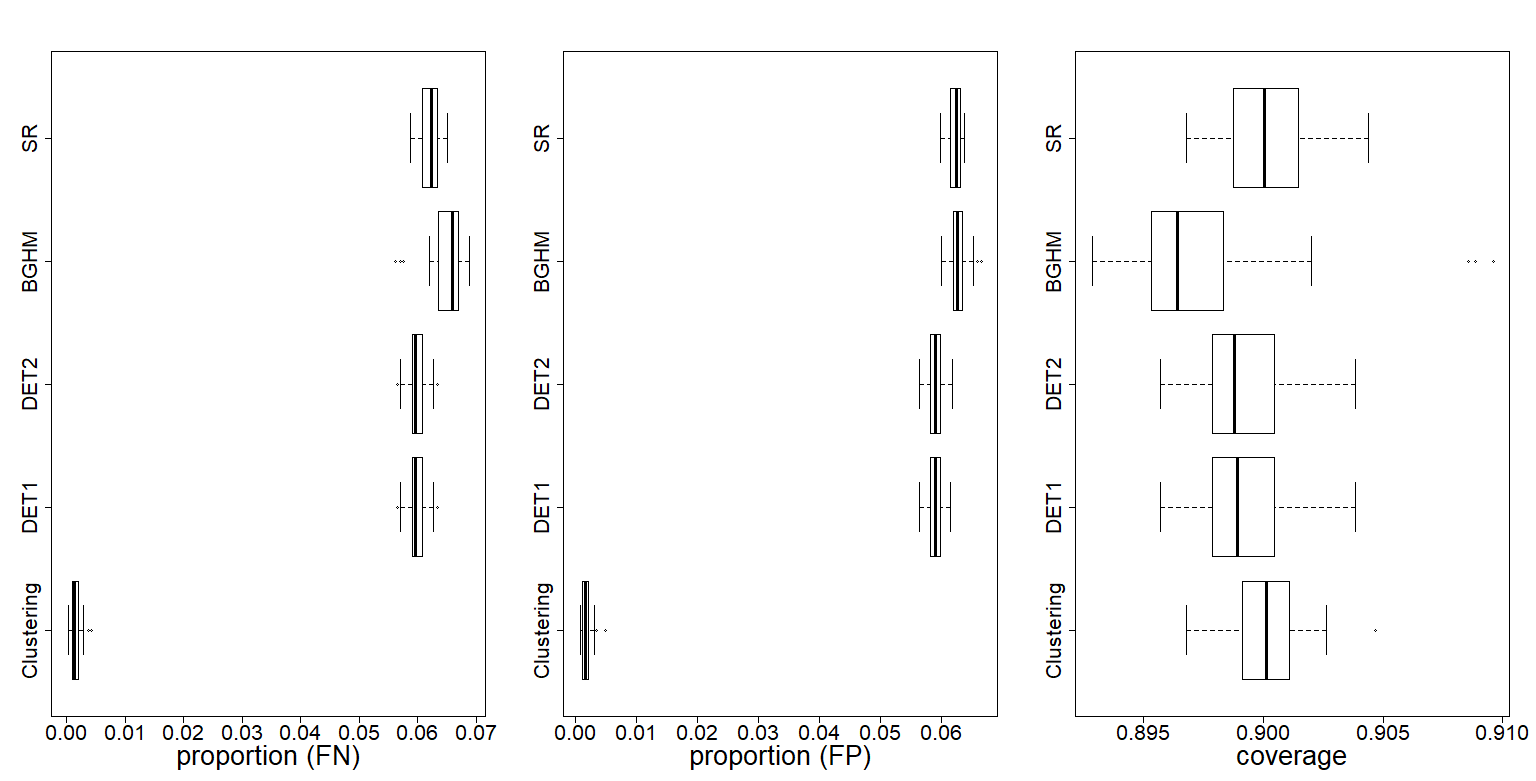}
   	\includegraphics[width=17cm,height=5cm]{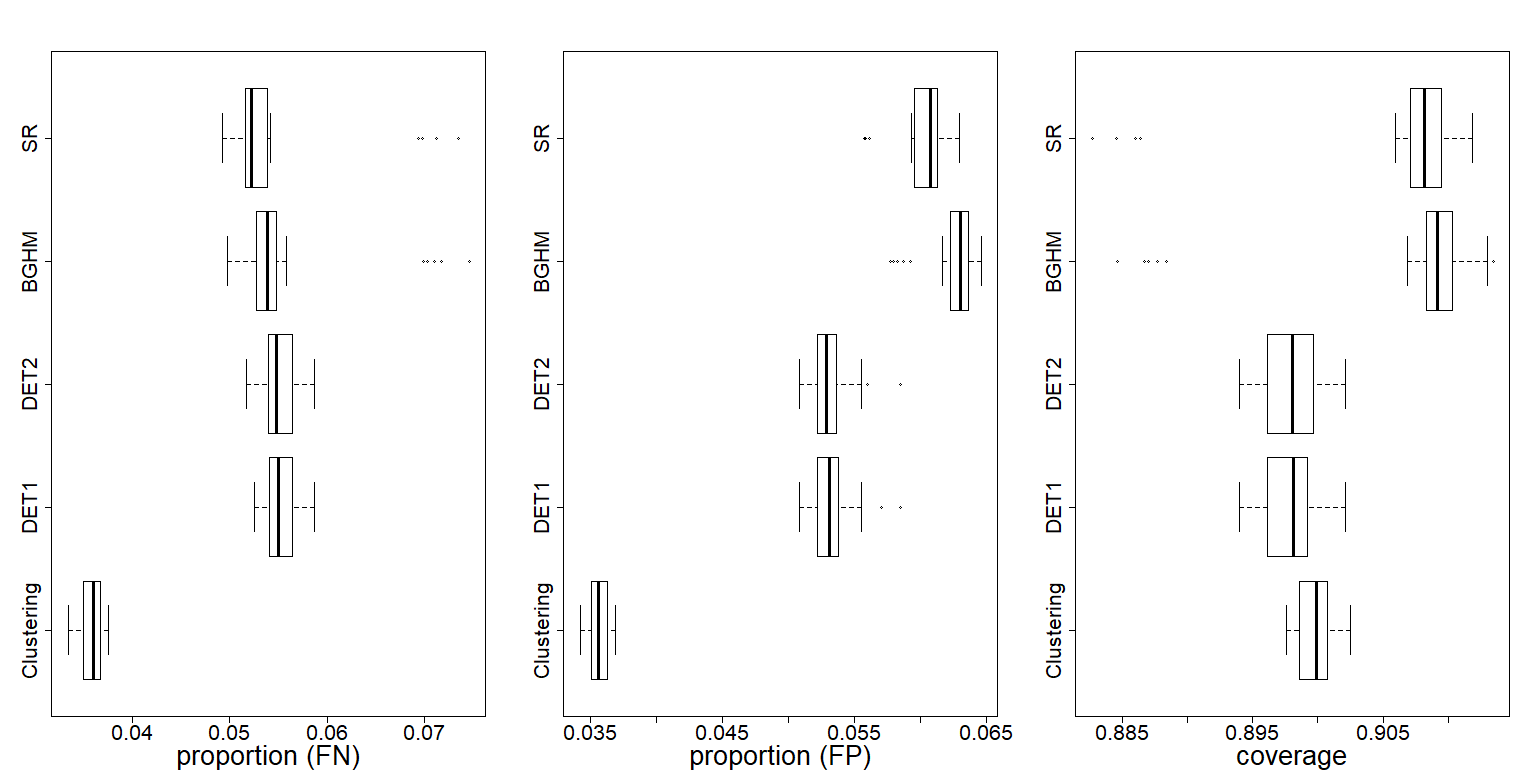}
   	\caption{Boxplot of false negative (left column, Eqn.~\ref{FN}), false positive (middle column, Eqn.~\ref{FP}) and coverage estimates $\hat\alpha$ of credible set estimations for the four distributions; (first row, i) banana shaped distribution, (second row, ii) donut shaped distribution, (third row, iii) skewed normal and (fourth row, iv) posterior distribution. The nominal coverage is $\alpha=0.9$.} \label{sample_fig}
   \end{figure}

	For each estimator, the false negative/positive rates and coverage of $Q^{KDE},Q^{Clust},\ldots,Q^{DET2}$ are computed using test samples $\pmb W^{(1:m')}$ ($m'=30,000$) independent of any samples $\Xb^{(1:n)}, \Zb^{(1:m)}$ used in the construction of the estimates. This is replicated 30 times. False negative and positive rates are estimated using the estimator in Theorem~\ref{thm5} with $\Zb^{(1:m)}$ replaced by $\pmb W^{(1:m')}$.
	Simulation results for a nominal coverage $\alpha=0.9$ are summarized in Figure~\ref{sample_fig}. The actual HPD set estimates for the two-dimensional banana (i) and donut (ii) examples are given in Figure~\ref{fig.hpd}. Although the coverage estimates plotted in Figure~\ref{sample_fig} are all close to the nominal value ($0.9$), the false negative/positive rates vary out to a maximum difference of $\pm 0.03$. The essentially univariate SR and BGHM methods are fast but these distributions are chosen to expose the assumptions of those methods and so although they match the coverage well, they have large loss values (ie, relatively large false positive and negative rates in columns one and two). This can be seen in Figure~\ref{fig.hpd} top left where the SR and BGHM set estimates, which are essentially square, cover white-space. The KDE method works well when it is practicable (first two rows) but is expensive to compute (and doesnt offer a convenient set representation) so it is omitted in the last two rows of plots in Figure~\ref{sample_fig} where model (iii,iv) dimension is higher. The parametric clustering method does poorly when its model assumptions are violated (concentric donuts (ii), second row) but is otherwise very effective. Finally, the two tree-based estimators, DET1 and DET2, which differ only in loss, have very similar performance, suggesting the strategy of minimising a lower bound on the loss
	(which is close to the actual loss when Theorem~\ref{thm6} applies) in DET1 is working well. In DET2 we have the advantage of being able to report an estimate of the loss. This is not possible in the settings in which we expect to use DET1. 
	
	The Gaussian mixture method Cluster is a simple practical alternative to DET1 and DET2 with good performance in Figure~\ref{sample_fig}. Table~\ref{tab_tree} shows density estimation tree sizes, numbers of Gaussian components used by Cluster and computing times (for DET1 and DET2, times are for DET2 as runtimes are similar). 
	For the clustering method, all 10 components are usually needed. Constructing a HPD set using the clustering method took 5 - 18 times longer than DET2 and so although we might get better performance measures by fitting models with larger numbers of clusters, computation time becomes prohibitive, at least in a straightforward implementation.
	
	\begin{table}[ht]  \centering \begin{tabular}{c|c|ccc|cc}
			
			&&DET1&DET2&DET2& \multicolumn{2}{c}{Cluster}  \\
			Distribution & dimension & \multicolumn{2}{c}{\# of subrectagles} & time (sec) & \# of clusters & time (sec) \\ \hline
			Banana shape (i) & 2 & 550 (12) & 355 (156) &370 & 9&2273 \\  \hline 
			Two donuts (ii) & 2 & 764 (21) & 762 (21) & 48 & 10 & 1024 \\  \hline
			Skew Normal (iii) & 10 & 3786 (99) & 3794 (91) &479 & 6&2863 \\ \hline
			Posterior (iv) & 14 & 3786 (99) & 3794 (91) &975&10&5969 \\ \hline
		\end{tabular}
		\caption{Runtime measures: target model dimension; number of sub-rectangles (leaves) of density estimation tree; mean value (standard error); time in seconds to produce a single final HPD set estimate (ie including all bandwidth selection and level-set estimation).} \label{tab_tree} \end{table}  
	
	Our conclusions from these studies are that DET1 and DET2 are useful robust algorithms with acceptable runtimes in a straightforward implementation. They produce explicit HPD set representations. 

	\subsubsection{Credible set estimation}\label{sec:real}
	
	In this section HPD-set estimates are given for two small real datasets using DET2 and a clustering based approach.
	
	\begin{description}
	\item[Radiocarbon-dating data]
	This nine dimensional example posterior is a slightly simplified version of the example considered in \citet{nicholls01}, Table~1, with data from \citet{anderson96}. The data set is available on  \url{http://www.radiocarbon.org/IntCal13%20files/shcal13.14c}. It is a fairly typical radiocarbon calibration problem in which the data are paired age and error measurements, respectively $y_i$ and $\sigma_i,\ i=1,\ldots,7$ and the observation model depends on known non-linear calibration and error functions $c(t), \sigma_c(t),\ t\ge 0$. The model has two parameters $\pmb \psi=(\psi_1,\psi_2)$ satisfying $L\le \psi_1<\psi_2\le U$ (with $L=500$ and $U=1000$) and seven date parameters $\pmb \theta=(\theta_1,\ldots,\theta_7)$ satisfying $\psi_1<\theta_i<\psi_2,\ i=1,...,7$ for $d=9$ parameters in all. Let $\psi^\pm=\psi_2\pm\psi_1$. The prior and observation model determining the posterior are as follows:
	\begin{align*}
	\psi^-&\sim U(0,U-L),\\
	\psi^+|\psi^-&\sim U(2L+\psi^-,2U-\psi^-),\\
	\theta_i&\sim U(\psi_1,\psi_2),\ &i=1,\ldots 7,\\
	y_i&\sim N(c(\theta_i),\sigma_c(\theta_i)^2+\sigma_i^2),\ &i=1,\ldots 7
	\end{align*}
	where data and parameters in the last two lines are both jointly independent.
	Posterior distributions $\pi(\pmb\psi,\pmb\theta|y)$ arising in radiocarbon calibration are often multimodal. Data set sizes are limited by budgets, so asymptotic normal approximations are not usually relevant. Also, the calibration function $c(t)$ \citep{hogg13} is one-to-many and so likelihood functions with multiple local maxima are common. In our example the joint distribution of $\psi_1,\psi_2|y$ is multimodal. The parameter space is bounded so there is no truncation in DET2.
	
	\item[Galaxy data]
	This is a benchmark example in mixture literature\footnote{\url{https://stat.ethz.ch/R-manual/R-devel/library/MASS/html/galaxies.html}}. Velocities $y_i, i=1,\ldots,82$ in km/sec of 82 galaxies from 6 well-separated conic sections of an unfilled survey of the Corona Borealis region are measured \citet{postman86}. We fit a three-component Gaussian mixture. The joint posterior is eight dimensional with six modes (due to label switching). For the Clustering method, a Gaussian mixture on logged weights is fitted. The observation model for each $y\in\mathbb R$ is 
	\[ p(y|\pmb \mu,\pmb\sigma) = \sum^3_{k=1} p_k N(y;\mu_k,\sigma_k^2),\qquad \sum^3_{k=1}p_k =1 \,. \]
    Conjugate priors are assigned on parameters, $(p_1,p_2,p_3)\sim Dir(1,1,1)$, $\mu_k\sim N(0,10^4)$ and $1/\sigma_k^2 \sim Gamma(3,3)$.
	\end{description}
	
\begin{figure}[!h] 
	\centering
	    \includegraphics[width=7cm]{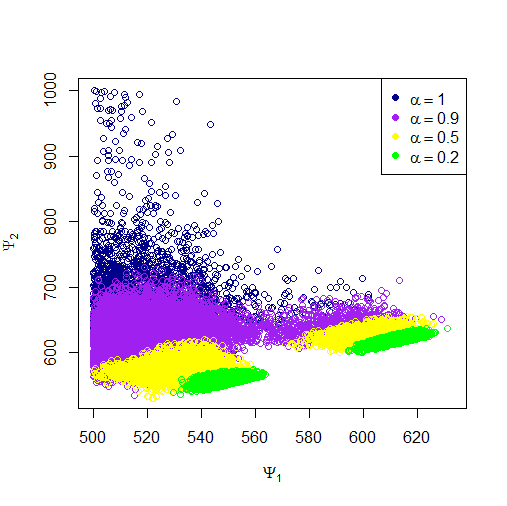} \includegraphics[width=7cm]{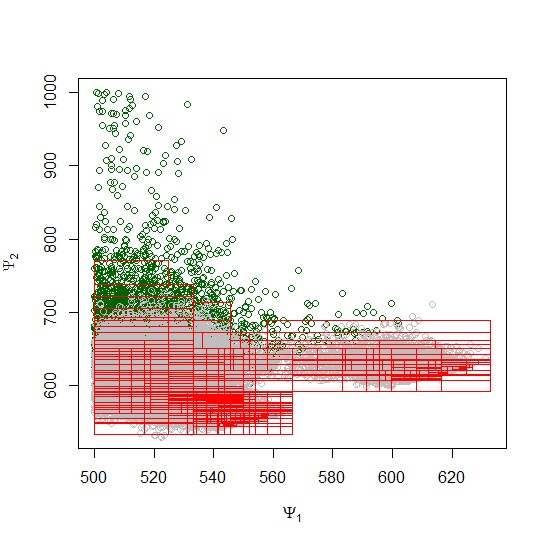}
	    \includegraphics[width=7cm]{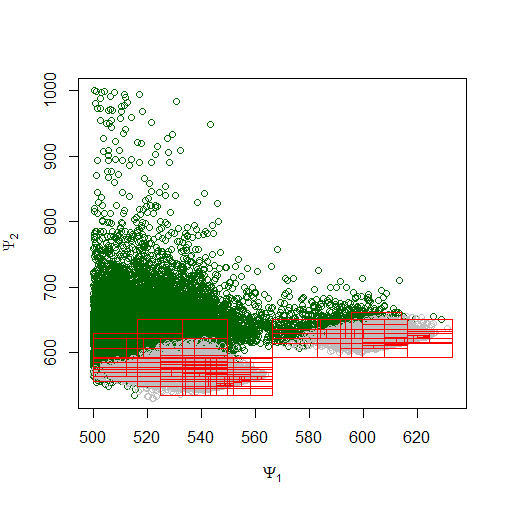}
	    \includegraphics[width=7cm]{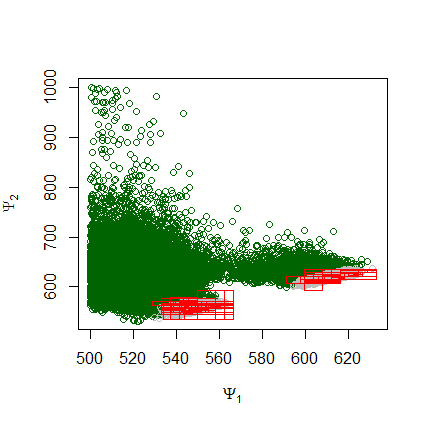}
\caption{300,000 posterior samples for $\Psi_2$ and $\Psi_1$; (Topleft) Points in HDP sets at the four levels ($\alpha=0.2,0.5,0.9,1.0$), Points inside/outside of a credible set are marked by grey/darkgreen and credible set estimation using the DET2 is marked by red boxes for the three levels $\alpha=0.9$ (Topright), $\alpha=0.5$ (Bottomleft) $\alpha=0.2$ (Bottomright). }\label{fig:rcd}
\end{figure}

	Performance measures for DET2 and Cluster are given in the Table \ref{tab:credibleset}. Fitting was done using the same set up as Section~\ref{sec:comparemethods} ($\tau$-value search, maximum number of clusters). The quality of the two credible set estimates are similar in terms of true/false positive rate, coverage and computing time. For the radiocarbon data, $s=\psi_2-\psi_1$ is an estimate of the time span of site occupation and is of particular interest. We plot the span $s$ against $\psi_2$ and the projection of the joint HPD set onto these parameters in Figure \ref{fig:rcd}. The two modes emerge as the target level is reduced. The projected HPD set top right includes white-space to the left and right. The space to the left of the diagonal would in practice be clipped at the known constraint $\psi_2+s\le U$. The apparently empty boxes to the lower right in this graph are large overlapping cells projected on top of one another in the projection from nine to two dimensions.

	\begin{table} \centering \begin{tabular}{c|c c|c c} 
	& \multicolumn{2}{c}{Radiocarbon-dating data} & \multicolumn{2}{c}{Galaxy data} \\ \hline 
	& DET2 & Clustering & DET2 & Clustering \\ \hline 
	FN & 0.040 & 0.027 & 0.036 & 0.090 \\
	FP & 0.039 & 0.022 & 0.032 & 0.091 \\
	Coverage & 0.89 & 0.89 & 0.90 & 0.90 \\
	Size & $K$=2907 & 10 clusters &$K$=1295 & 10 clusters \\ 
	Time (Sec) & 667 & 2409 & 391 & 882 \\ \hline 
	\end{tabular} \caption{True/False positive rates, coverage (nominal value is $\alpha=0.9$) for credible set estimation, tree sizes and computing time in seconds.} \label{tab:credibleset} \end{table}  

\subsubsection{HPD sets for parameter functions}\label{sec:postpred}

It may be of interest to construct Highest Posterior credible sets for functions of parameters.
The {\tt hsbdemo} data set\footnote{\url{http://www.ats.ucla.edu/stat/data/hsbdemo.dta.}} contains 200 high school student's performance in USA and their program choices among general program, vocational program and academic program. Thirteen predictor variables include gender, social economic status and various test scores. 

A multinomial regression is used to model program choices. Given a vector of predictors $r_i$ for student $i=1,\ldots,200$, the program choice $y_i\in\mathcal{Y}_0,\ \mathcal{Y}_0=\{``general",``academic",``vocational"\}$ is distributed as $y_i \sim Multinomial(\nu_i),\ \nu_i=(\nu_{i,j})_{j\in \mathcal{Y}_0}$ with  $\sum_{j\in \mathcal{Y}_0}\nu_{i,j}=1$. Program probabilities $\nu_{i,j}=\nu(r_i,\beta_j)$ where 
\[
\nu(r_i,\beta_j) = \frac{exp(r_{i}\beta_j)}{\sum_{k\in \mathcal{Y}_0} exp(r^T_{i}\beta_k)}
\] 
with $\beta_j\in \mathbb{R}^D$ for $j\in \mathcal{Y}_0$.

Our predictor variables are a gender ($g=$ "female" or "male"), social economic status ($ses=$ ``low" or ``medium", or ``high") and writing score ($w$) so that $D=4$. An independent improper uniform prior is assigned for all elements of $\pmb \beta=(\beta_j)_{j\in\mathcal{Y}_0}$ with $\pmb\beta\in \mathbb{R}^{12}$. Posterior samples are generated using the R-package \texttt{MCMCpack}. If the posterior is $\pi(\pmb\beta|\yobs)$ when $\yobs=(y_1,\ldots,y_{200})$ then we are interested in the posterior distribution of $\nu(\pmb\beta_j;r),\ j\in \mathcal{Y}_0$ for $r$ some new student's covariate vector. The program probabilities $\nu(\pmb\beta_j;r)$ with covariates $r$ determined for a female student with $ses=$ ``high" and a writing score of 36 are predicted using posterior samples of $\pmb \beta\sim \pi(\cdot|\yobs)$. 

\begin{figure}[!h] \centering
\includegraphics[width=7cm]{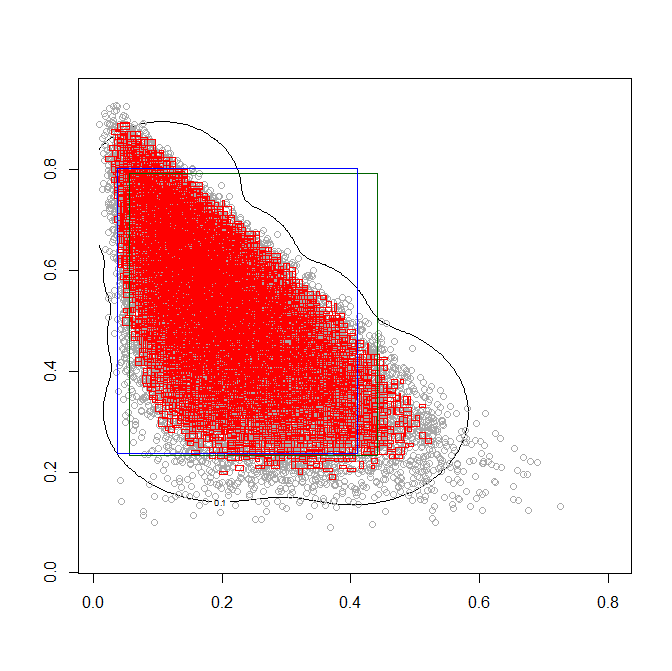}  
\includegraphics[width=7cm]{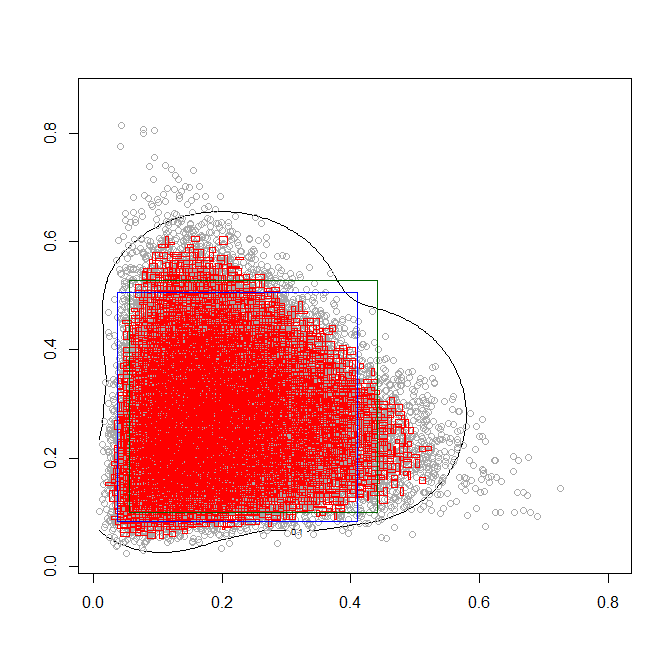}
\caption{30,000 samples for (Left) $\nu_{general}
$-vs-$\nu_{vocational}$ and (Right) $\nu_{general}
$-vs-$\nu_{academic}$ are marked by grey. Credible set estimation using the DET1 (red boxes), SR (blue box), BGHM (dark green box) and a clustering method (black line).}\label{fig:ml} 
\end{figure}

Since the three components of $\nu$ sum to one, the 90\% credible set is a set in two dimensions. It is graphically presented in Figure~\ref{fig:ml} for the Cluster, SR, BGHM and DET1 set estimators of Section~\ref{sec:comparemethods}. The two graphs show HPD sets for pairs of posterior functions $\nu(\pmb\beta_j,r)$ and $j=``general",``vocational"$ (left) and $j=``general",``academic"$ (right). In this example we have not calculated the target density up to a constant, so we use the DET1 estimator and cannot give loss estimates. Visual inspection suggests Cluster, SR and BGHM cover too much whitespace. It seems the Cluster EM algorithm has not converged to an adequate approximation.
On the other hand it can be seen that the DET1 set is the union of many small sets and is fragmented at the edge. In a higher dimensional setting this would make estimation of set-topology unreliable.
The Posterior probability for this individual tends to be higher for the vocational program and lower for general program. However, it can be seen in Figure \ref{fig:ml} with at least some small probability, different orders in probabilities are supported by the data. 

	\subsection{Bayesian calibration}\label{sec:calibexample}
    
    Variational Bayes and Approximate Bayesian Computation (ABC) methods are popular approximation methods in Bayesian inference and machine learning \citep{beaumont02, blum10, ormerod10, jordan99}. When we estimate HPD sets using these approximations form an HPD set for a density which only approximates the true posterior. We calibrate these ``doubly approximate'' HPD set estimates using the method of Section~\ref{sec:calibtheory}. In the examples which follow the nominal coverage is $\alpha=0.9$, $n=3\times 10^5$, $m=3\times 10^4$ and the coverage estimate is given using $\hat c(\yobs)$ as explained in Section~\ref{sec:calibtheory}.
    
\subsubsection{Exchange rate data}\label{sec:gk}

Total 1985 daily NZ dollar exchange rate against US Dollars\footnote{\url{https://www.ofx.com/en-nz/forex-news/historical-exchange-rates/}} $y'_t, t=1,...,T+1$ for $T=1984$ days from 23 March 2017 to 21 March 2020 are transformed to scaled log returns $y_t=500\times \log(y'_{t+1}/y'_{t})$. The log daily return distribution is modelled by the $g$-and-$k$ distribution model for highly skewed or heavy tailed data. The distribution is defined by transforming a standard normal random variable $z\sim N(0,1)$ to $ y = A+BG(z)H(z) $ with the asymmetry and elongated tails expressed using $G(z)=1+c\tanh(gz/2)$ and $H(z)=z(1+z^2)^k$ respectively. This g-and-k distributions uses a quantile function of the form $F^{-1}(u;A,B,g,k,c)=Q_{gk}(\Phi^{-1}(u);A,B,g,k,c)$ where $\Phi(\cdot)$ is the $N(0,1)$ CDF and 
\[ Q_{gk}(z;A,B,g,k,c) = A+B(1+c\tanh(gz/2)) z(1+z^2)^k \,, \]
It is standard to take $B>0$ and $c=0.8$ to guarantee a valid distribution. Uniform priors are assigned for the four parameters \[ A\sim U(-1,1)\,, B\sim U(0,1)\,, g\sim U(-5,5)\,, k\sim U(0,10)  \,. \]
The generative model for this doubly intractable posterior can be simulated. The posterior itself can also be simulated, and these features have made it a popular example in the ABC literature. 


The ABC-MCMC method of \citet{marjoram13} and the extension by \citet{Wegmann09} were implemented using the seven order statistics of the data $y$ as the ABC summary statistics. The $d=4$ dimensional exact posterior distribution and ABC-type approximations are simulated using the R-packages \texttt{gk} and \texttt{EasyABC} respectively. Credible sets of ABC approximations are estimated using DET1 as the ABC posterior density is not available.

\begin{figure}[!h] 
	\centering
	\includegraphics[width=12cm,height=8cm]{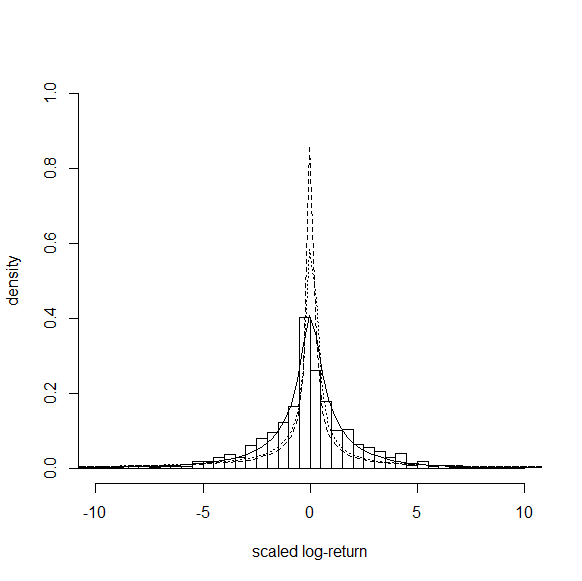}
	\caption{Histogram of scaled-log return data and posterior predictive density $q_{gk}$ using exact MCMC simulation (solid), ABC by \citet{marjoram13} (dashed) and ABC by \citet{Wegmann09} (dotted line).}  \label{exch_fig}
\end{figure}

The posterior predictive density $q_{gk}(z)$ of the quantile function is given in Figure~\ref{exch_fig}, estimated using exact posterior samples and ABC samples. The ABC-methods yield tighter densities around 0 and have longer tails compared to the exact posterior expectation. We estimate joint four-dimensional HPD sets for the two approximate posteriors using DET1. These have coverage close to $0.9$ in the approximate posterior. However their coverage in the true posterior is quite different. The {\it true} coverage of the credible set approximations due to \citet{marjoram13} and \citet{Wegmann09} are 0.978 and 1 respectively (estimated to stated precision using exact posterior samples). These sorts of estimates (ie based on samples from the true doubly intractable target) are not usually available so it is desirable to estimate true coverage in the indirect methods of \cite{Xing2019}. Using the BART and logistic regression, we estimate the true coverage values as $0.968 \pm 0.053$ for ABC-MCMC method by \citet{marjoram13} and $0.972 \pm 0.033 $ for ABC method by \citet{Wegmann09}. This is useful reassurance. The coverage of the estimated HPD sets using the posterior-approximations are too high, and calibration allows us to measure this.

\subsubsection{West Nile virus data}\label{sec:wnv}

West Nile virus is most commonly spread to humans through infected mosquitoes. The City of Chicago and the Chicago Department of Public Health (CDPH) set mosquito traps across the city and every week from late spring through the fall, they are tested for the virus. The results of these tests influence when and where the city will spray airborne pesticides to control adult mosquito populations. The dataset is available in \url{https://www.kaggle.com/c/predict-west-nile-virus}.

\begin{figure} \centering
\includegraphics[width=6cm,height=6cm]{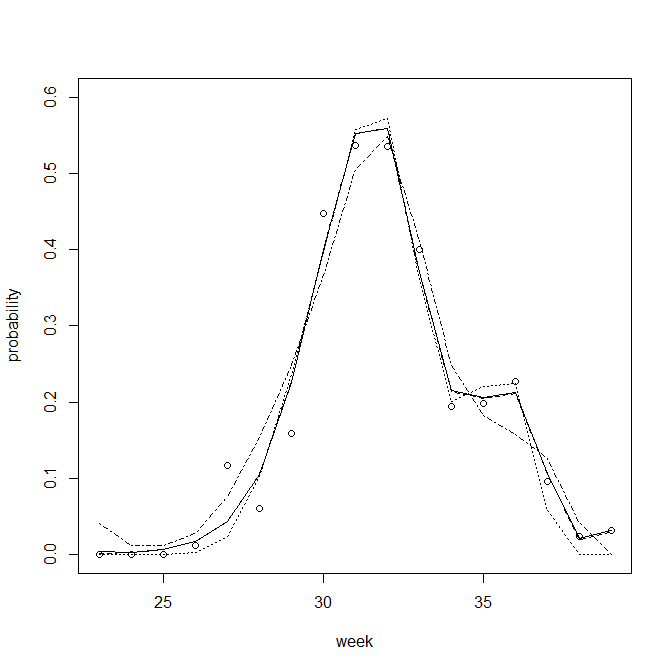} 
\includegraphics[width=6cm,height=6cm]{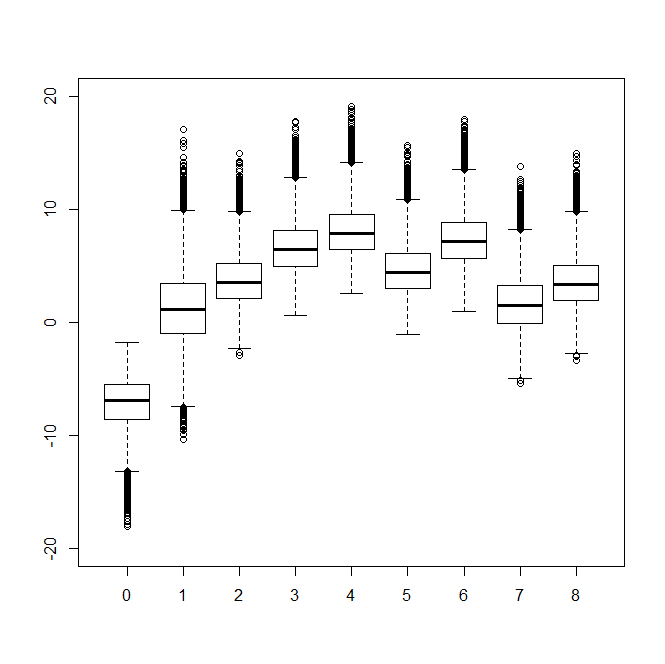}
\includegraphics[width=6cm,height=6cm]{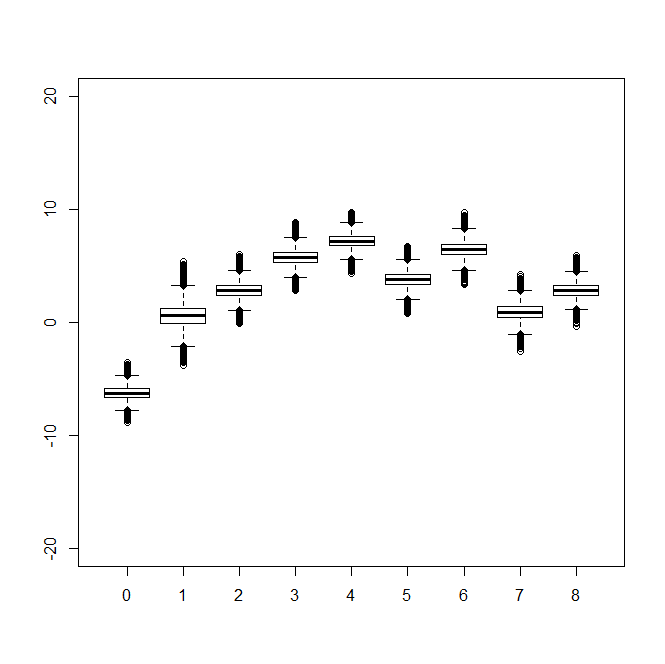}
\includegraphics[width=6cm,height=6cm]{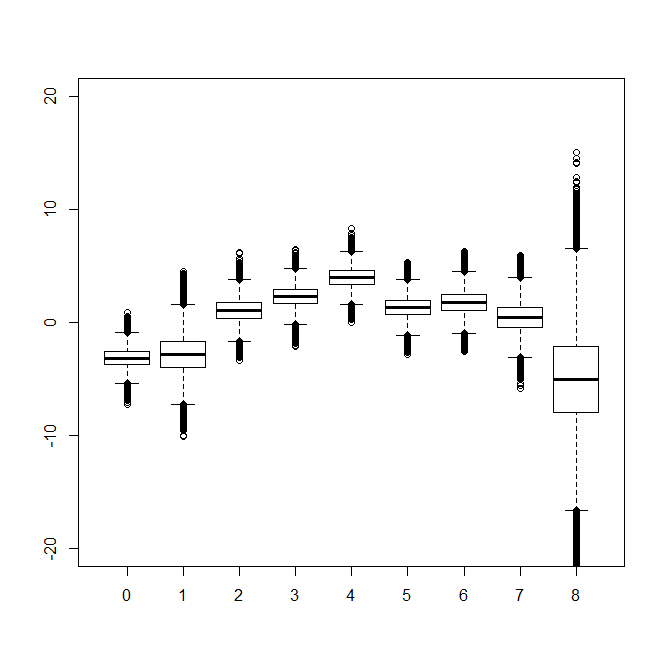}
\caption{Top-Left: Observed probability of positive rest result (dots) and posterior expectations using the exact posterior samples (solid line), using CAVI (dashed line), SVI (dotted line) and Laplace approximation (dash-dot line). Top-Right: Exact marginal posterior distributions. Bottom-Left: Marginal posterior distributions due to CAVI. Bottom-Right: Marginal posterior approximate due to Laplace approximation.} \label{westnile}
\end{figure}

The probability for a positive test result with week numbers in year 2016 is modelled using logistic regression. There are $T=2029$ test results $y_i\in \{0,1\},\ i=1,\ldots,T$ from Weeks 23 to 39. The proportion of positive results per week is shown in Figure \ref{westnile}. A B-Spline regression with 5 knots is fitted to the test success probability as a function of time yielding a $d=9$ dimensional joint posterior distribution with spline parameters $B\in \mathbb{R}^d$. 

	Classical Bayesian logistic regression for $y_i$ given a matched vector $\pmb w_i$ of $d$ spline covariate functions of the recording time is 
	\[ y_i|B \sim {\rm Bern} \left[ \frac{e^{ (1, \pmb w_i)^T B}}{1+e^{(1, \pmb w_i)^T B} } \right] \,, i=1,...,{n_L} \]
	and a normal prior is assigned for coefficients $B$; $B \sim N(\mu_B,\Sigma_B)$ where $\mu_B=[0,...,0]$ and $\Sigma_B=100I_{(d+1)\times (d+1)}$. The exact posterior distribution is simulated using the R-package, \texttt{brms}. Two readily available posterior approximations on which we can base HPD set estimates are Variational Bayes and Laplace approximation. Recently \citet{Durante2019} showed the connection between the P\'{o}lia-gamma data augmentation and the variational approach \citep{Jaakkola2000} for logistic regression. Their simulation code for stochastic variational inference (SVI) and coordinate ascent variational inference (CAVI) are available on \url{https://github.com/tommasorigon/logisticVB}. Credible sets of posterior approximations are estimated using DET2 as posterior approximation densities are available for variational methods.

Posterior expectations for the positive result probability due to variational Bayes approaches and Laplace approximation capture the trend of the test results well in Figure~\ref{westnile} (top left) but disagree in distribution quite significantly (other three panels). We visualise these using scalar marginal distributions as the difference is already clear. The operational coverage $c(\yobs)$ of the credible set approximations may be estimated directly from the posterior samples and is 0.181 for the CAVI, 0.197 for SVI and 0.028 for Laplace approximation, differing substantially from the nominal level of $0.9$.  

\section{Conclusion}\label{sec:conclusion}


We have given an estimator for joint HPD sets for mutivariate densities based on samples from the density. The difference between the joint HPD set and an HPD set formed from a product of marginals, as usually used, can be substantial. If a marginal-product HPD set is to be used, we suggest it can at least be checked by estimating the posterior mass it puts on the symmetric set difference with the true HPD set (and this can be done without joint HPD-set estimation). Algorithms~\ref{alg:DI} and \ref{alg:tractablepost} compute the HPD set estimate in the case where one respectively cannot and can compute the target density up to a constant. The first case is important as doubly intractable posterior distributions arise in Bayesian inference. In the second case one can estimate the probability mass on the symmetric difference between the estimated set and the truth. We use this to tune the bandwidth in Algorithm~\ref{alg:tractablepost}. In practice we found Algorithm~\ref{alg:DI}, which is based on minimising a lower bound on the loss, performs about as well without this additional information, presumably because the tree-estimator is doing its job well, and by forming the set-estimate from the highest density leaves we really are searching over HPD-sets as we vary $\tau$ and $\alpha_{n,\tau}$.

KDE-based set-estimation is statistically efficient in all cases but rather computationally demanding to compute on large data sets so although we included KDE HPD sets in the comparison studies in low dimensional examples we dropped it from comparison in higher dimensions. We found that DET-based joint HPD set estimation is competitive with a method (``Cluster'') based on fitting a mixture of up to ten multivariate normal densities using EM and taking the level set. Our runtime comparison is indicative only - but DET based HPD-set estimation is not slower. The quality of the estimated HPD sets as measured by loss is similar, with Cluster doing better when its parametric assumptions were well matched by the target and worse if not. The memory requirement for representing the HPD set from a DET is $O(Kd)$ and $O(J_{Clust}d)$ for the clustering method. From our experience $K$ tends to be much larger than  $J_{Clust}$ but much smaller than $n$.
 
HPD sets are useful for learning about the structure of the posterior. DET-based HPD sets have a simple geometric representation in terms of hyper-rectangular leaf sets. However, further work is needed to summarise and visualise high-dimensional sets. The DET framework of binary splits offers a convenient description to explore the structure of level-sets and compute similar summary-graphs to those developed in \citet{chen17} and further topological summary statistics may be of interest. We have not explored this in any detail. When the topology of the level set is of interest, a set shape-constraint \citep{baillo06} may need to be added to the loss, as a loss based on the measure of symmetric set-difference allows set estimates that capture the bulk of the distribution but include ``fragments'' at the edges.

The calibration procedure we use allows us to estimate the coverage of a multidimensional set in the true posterior without having samples from the posterior. This is done using logistic regression in data space. The estimated success probability function (over the data) is the coverage when evaluated 
at the data.
This is straightforward when the dimension of data-space is not too large, as is the case when there are sufficient statistics in doubly intractable exponential family models.
The method could be used to calibrate HPD sets computed on an approximation at multiple levels $\alpha_1,\alpha_2,\ldots$ (the cost is one logistic regression at each $\alpha$-value) and investigate how well level sets match between truth and approximation. 

\section*{Appendix}

The density estimation tree algorithm of \citet{Li2016} and the maximum gap calculation it uses are given in this appendix.

\subsubsection*{Maximum gap calculation}

In order to find a good split for leaf $\Delta_k=[\pmb a^{(k)},\pmb b^{(k)}]$, defined in Step~5 of Algorithm~\ref{alg:detcore}, and given a set of points $\{\wsb^{(k,j)}\}_{j=1}^{n_k}$ with $\wsb^{(k,j)}=(\tilde s^{(k,j)}_{1},...,\tilde s^{(k,j)}_{d})$, we divide the $i$-th dimension into $m_g$ equal-sized bins, $[a_{k,i}+(l-1)\delta_{k,i},a_{k,i}+l\delta_{k,i}]$, $i=1,...,d$ and $l=1,...,m_g$ where $\delta_{k,i}=(b_{k,i}-a_{k,i})/m_g$. There are in total $(m_g-1)d$ gaps. Each gap is defined by $h_{l,i}=| (1/n_k) \sum^{n_k}_{j=1}  \mathbbm{1}( \tilde s^{(k,j)}_i < a_{k,i}+l\delta_{k,i}) -l/m_g | $, $l=1,...,(m_g-1)$ and $i=1,...,d$. The splitting hyperplane is the gap with the maximum $h$-value.


\begin{algorithm}
	\caption{Density estimation tree}
	Input: Dataset $\xb^{(1:n)}$, ${\Omega}^{(p)}=[{\omega}^{p,-},{\omega}^{p,+}]$ and $\tau$.
	
	Output: A piecewise constant $\widetilde{f}_n^{(p)}$ defined on a binary partition $\Lambda$. 
	
	\begin{algorithmic}[1]\label{alg:detcore}
	
	\STATE Sort $\xb^{(1:n)}$ so that $\xb^{(1:N)} \in {\Omega}^{(p)}$ are the $N$ points in ${\Omega}^{(p)}$ and discard $\xb^{(N+1:n)}$. 
	
	\STATE Transform $\xb^{(j)}$ to $\sb^{(j)} \in [0,1]^d$, $j=1,...,N$ using the linear transformation,
    \[
    s_i^{(j)}=\frac{X_i^{(j)}- {\omega}_i^{p,+}}{{\omega}_i^{p,+}-{\omega}_i^{p,-}},\quad i=1,...,d.
    \] 
    Set $\Delta_1=[\ab^{(1)},\bb^{(1)}]$ where $\ab^{(1)}=(0,0,\ldots,0)$ and $\bb^{(1)}=(1,1,\ldots,1)$ are $d$-component vectors, $\Delta=\{\Delta_1\}$ 
    and $n_1=N$. Set Process={\it incomplete}.
		\WHILE {Process={\it incomplete}}
		\STATE $\Delta'=\emptyset$.
		\FOR{each $\Delta_k=[\pmb a^{(k)},\pmb b^{(k)}]$ in $\Delta$}
		\STATE Denote points in $\Delta_k$ by $\{ \sb^{(k,j)} \}_{j=1}^{n_k}$. Using cell boundaries $\ab^{(k)}=(a_{k,1},...,a_{k,d})$ and $\bb^{(k)}=(b_{k,1},...,b_{k,d})$, rescale with $\wsb^{(k,j)}=\left ( \frac{s^{(k,j)}_1-a_{k,1}}{b_{k,1}-a_{k,1}},...,\frac{s^{(k,j)}_d-a_{k,d}}{b_{k,d}-a_{k,d}} \right )$, $j=1,...,n_k$.
		\STATE Calculate gaps $\{ h_{l,i} \}_{i=1,...,d, l=1,...,m_g-1}$ in $\Delta_k$ and $D^*(\{\wsb^{(k,j)}\}_{j=1}^{n_k})$ (see Appendix).
		\IF{$n_k>2$ and $D^*(\{\wsb^{(k,j)}\}_{j=1}^{n_k} )>\tau\sqrt{N}/n_k$ }
		\STATE Split $\Delta_k$ into $\Delta_{k1}=[\pmb a^{(k1)},\pmb b^{(k1)}]$ and $\Delta_{k2}=[\pmb a^{(k2)},\pmb b^{(k2)}]$ along the max gap. 
		\STATE Compute $n_{k1}=\sum_{j=1}^{n_k} \mathbbm{1}(\wsb^{(k,j)}\in \Delta_{k1})$ and $n_{k2}=n_k-n_{k1}$. 
		\STATE $\Delta'=\Delta'\cup \{ \Delta_{k1},\Delta_{k2} \}$
		\ELSE 
		\STATE $\Delta'=\Delta'\cup \{ \Delta_{k}\}$
		\ENDIF 
		\ENDFOR
		\IF{$\Delta'\neq \Delta$} \STATE $\Delta=\Delta'$ 
		\ELSE 
		\STATE Process={\it complete}
	    \ENDIF 
	    \ENDWHILE 
	    \STATE Transform partition $\Delta$ back to a partition of $\Omega^{(p)}$. Set $\Lambda_k=[\ub^{(k)} ,\vb^{(k)}]$ with $\ub^{(k)}=(u_{k,1},...,u_{k,d})$ and $\vb^{(k)}=(v_{k,1},...,v_{k,d})$ where $u^{(k)}_i= a^{(k)}_i({\omega}_i^{p,+}-{\omega}_i^{p,-})+{\omega}_n^{p,-}$ and $v^{(k)}_i= b^{(k)}_i({\omega}_j^{p,+}-{\omega}_j^{p,-})+{\omega}_n^{p,-}$, $i=1,...,d$ and $k=1,...,K$. 
	    The piecewise constant density for $\xb$ is 
	    \[
	    \widetilde{f}^{(p)}_n(\xb)=\sum^K_{k=1}\frac{n_k/N}{|\Lambda_k|}\mathbbm{1}(\xb\in \Lambda_k)
	    \]
	    and $\Lambda=\{\Lambda_1,\ldots,\Lambda_K\}$.
	\end{algorithmic}
\end{algorithm}

\clearpage
\bibliographystyle{chicago}  
\bibliography{reference}

\end{document}